\documentclass[journal=ancac3,manuscript=article]{achemso}

\usepackage{amssymb}
\usepackage{graphicx}
\usepackage[T1]{fontenc} 
\usepackage{chemformula} 
\usepackage{comment}
\AtBeginDocument{\singlespacing}

\author{Hannah L. Weaver}
\affiliation[University of California, Berkeley]{Department of Physics, University of California, Berkeley, CA 94720, United States}

\author{Cora M. Went}
\affiliation[California Institute of Technology]{Department of Physics, California Institute of Technology, Pasadena, CA 91125, United States}

\author{Joeson Wong}
\affiliation[California Institute of Technology]{Department of Applied Physics and Materials Science, California Institute of Technology, Pasadena, CA 91125, United States}
\altaffiliation{Current address: Department of Chemistry, University of Chicago, Chicago, IL 60637, United States}

\author{Dipti Jasrasaria}
\affiliation[University of California, Berkeley]{Department of Chemistry, University of California, Berkeley, CA 94720, United States}
\altaffiliation{Current address: Department of Chemistry, Columbia University, New York, NY 10027, United States}

\author{Eran Rabani}
\affiliation[University of California, Berkeley]{Department of Chemistry, University of California, Berkeley, CA 94720, United States}
\alsoaffiliation{Materials Sciences Division, Lawrence Berkeley National Laboratory, Berkeley, CA 94720, United States}
\alsoaffiliation{The Sackler Center for Computational Molecular and Materials Science, Tel Aviv University, Tel Aviv, Israel 69978}

\author{Harry A. Atwater}
\affiliation[California Institute of Technology]{Department of Applied Physics and Materials Science, California Institute of Technology, Pasadena, CA 91125, United States}

\author{Naomi S. Ginsberg}
\email{nsginsberg@berkeley.edu}
\affiliation[University of California, Berkeley]{Department of Chemistry, University of California, Berkeley, CA 94720, United States}
\alsoaffiliation{Department of Physics, University of California, Berkeley, CA 94720, United States}
\alsoaffiliation{Molecular Biophysics and Integrated Bioimaging Division, Lawrence Berkeley National Laboratory, Berkeley, CA 94720, United States}
\alsoaffiliation{Materials Sciences Division, Lawrence Berkeley National Laboratory, Berkeley, CA 94720, United States}
\alsoaffiliation{Kavli Energy NanoScience Institute, Berkeley, CA 94720, United States}
\alsoaffiliation{STROBE, NSF Science \& Technology Center, Berkeley, California 94720, United States}

\title{Detecting, distinguishing, and spatiotemporally tracking photogenerated charge and heat at the nanoscale}

\keywords{excitons, thermometry, energy transport, optical properties, nanoscale}

\begin{document}

\pagebreak
\begin{tocentry}

\includegraphics[width=7.5cm]{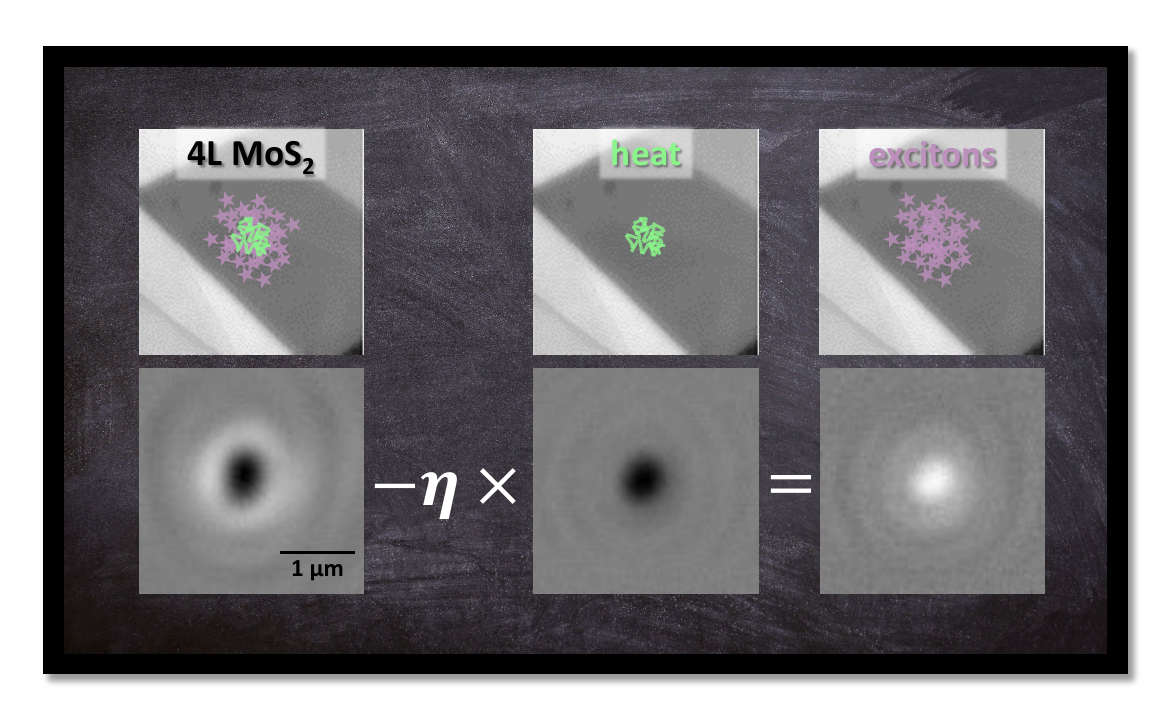}

\end{tocentry}

\begin{abstract}

Since dissipative processes are ubiquitous in semiconductors, characterizing how electronic and thermal energy transduce and transport at the nanoscale is vital for understanding and leveraging their fundamental properties. For example, in low-dimensional transition metal dichalcogenides (TMDCs), excess heat generation upon photoexcitation is difficult to avoid since even with modest injected exciton densities, exciton-exciton annihilation still occurs. Both heat and photoexcited electronic species imprint transient changes in the optical response of a semiconductor, yet the unique signatures of each are difficult to disentangle in typical spectra due to overlapping resonances. In response, we employ stroboscopic optical scattering microscopy (stroboSCAT) to simultaneously map both heat and exciton populations in few-layer \ch{MoS2} on relevant nanometer and picosecond length- and time scales and with 100-mK temperature sensitivity. We discern excitonic contributions to the signal from heat by combining observations close to and far from exciton resonances, characterizing photoinduced dynamics for each. Our approach is general and can be applied to any electronic material, including thermoelectrics, where heat and electronic observables spatially interplay, and lays the groundwork for direct and quantitative discernment of different types of coexisting energy without recourse to complex models or underlying assumptions.

\end{abstract}

\section{}

Heat and charge coexist in many semiconducting materials following photoexcitation or electrical injection. Here ``heat'' refers to heating induced lattice fluctuations, and ``charge'' refers to excited state electrons, holes and their correlated combinations, e.g., excitons. In low-dimensional materials, Auger-Meitner (A-M) processes, density-dependent heat-generating exciton annihilation events, are prevalent even at modest exciton densities due to enhanced Coulomb interactions, \cite{wang_auger_2006} and nonradiative pathways often dominate due to defects and natural background doping. \cite{lien_electrical_2019, uddin_enhanced_2022} The combination of these effects can lead to a significant fraction of absorbed light energy undergoing transduction to lattice heating, which coexists with other charge excitations like excitons. Similarly, in nanoscale electronic devices, charge carrier scattering with phonons leads to Joule heating and elevated device temperatures that impair efficient electronic dynamics, including transport, due to increased scattering with the lattice \cite{das_transistors_2021}. As device dimensions and volume for heat dissipation continue to decrease, material interfaces increase, and carrier-boundary scattering also plays a key role in self-heating, limiting thermal and electrical conductivity \cite{yalon_temperature-dependent_2017}. Each of these dissipative effects in optoelectronic devices is important to discern so that metrics such as photoluminescent quantum yield (PLQY) and carrier diffusion length can be optimized. Additionally, thermal management strategies often leverage inherent material anisotropies in energy flow which give rise to thermoelectric capabilities, the ability to reversibly convert an electric potential, $V$, to a temperature gradient, $\nabla T$, or vice versa. In either case, distinguishing heat from charge when they coexist in a material and discerning their unique photoinduced dynamics is vital not only for informing design principles for directing heat and charge in emerging materials but also for drawing well-informed conclusions about intrinsic material properties.

Distinguishing between heat and charge with optical measurements is, however, challenging. For example, both ground state and transient excited state optical spectroscopy manifest complex perturbations to the location, amplitude, and width of electronic resonances. \cite{varshni_temperature_1967, cooper_physical_2018, ruppert_role_2017} Even if signatures of unique excitations are quantitatively observed, because their signatures can overlap spectrally and temporally, they remain difficult to quantitate despite judicious choices in excitation and probing wavelengths, time-dependent signatures, measuring as a function of voltage bias, and measuring a pump fluence- or temperature dependence. In particular, local heating could influence or masquerade as electronic excitations in semiconducting materials, and because it is spectrally ubiquitous, isolating the electronic dynamics is especially challenging. Physically, heating leads to increased lattice fluctuations, which change the mass density, an effect measured by the thermo-optic coefficient and which may be observed as a frequency-independent change in the dielectric function. \cite{waxler_effect_1973, fan_thermal_1998} Near an electronic resonance, however, heating broadens and shifts the resonance feature, an effect that adds to or cancels any transient photoinduced changes from the electronic carriers themselves. \cite{ruppert_role_2017, cooper_physical_2018} This behavior is especially complicated in transition metal dichalcogenides (TMDCs) where heat dissipation following photoexcitation is prevalent and different exciton resonance features may spectrally overlap. \cite{pogna_photo-induced_2016}

The growing collection of spatiotemporally resolved optical microscopies are excellent candidates for characterizing photogenerated energy and its transport in semiconducting materials because they provide spatial dynamics in addition to the information provided by more traditional time-resolved spectroscopy. These techniques, e.g., micro-time-resolved photoluminescence (microTRPL), \cite{penwell_resolving_2017, kulig_exciton_2018, zipfel_exciton_2020, goodman_substrate-dependent_2020} transient absorption microscopy (TAM), \cite{shi_exciton_2013, gabriel_direct_2013, cui_transient_2014}, variations of transient reflectance microscopy, \cite{grumstrup_pumpprobe_2015, hill_perovskite_2018, block_tracking_2019} including stroboscopic scattering microscopy (stroboSCAT), \cite{delor_imaging_2020}   directly measure the progressive expansion of initially localized populations of impulsively photogenerated energy carriers with nanoscale sensitivity and down to ultrafast time resolution. TRPL is, however, restricted to detecting electron-hole radiative recombination, and, to our knowledge, there are no reports of TAM separately resolving heat directly in addition to charge. Fortunately, due to its high sensitivity to changes in the real part of the dielectric function, even beyond transient reflectance alone, stroboSCAT has investigated heat flow in metallic composite films \cite{guzelturk_nonequilibrium_2020, utterback_nanoscale_2021} and has distinguished heat and charge flow in p-doped silicon based on different signs of imaging contrast, a substantial separation in time scales, and corroboration with commonly cited respective diffusivities. \cite{delor_imaging_2020} To address the additional challenges associated with characterizing electronic dynamics and transport in ultrathin semiconductors, especially when heat and electronic diffusivities are not as dissimilar as in silicon and where photogenerated heat is prominent, additional strategies must be developed.

Here, we directly co-measure both heat and excitons in few-layer molybdenum disulfide (\ch{MoS2}) on relevant nanometer and picosecond length- and time scales using stroboSCAT. Through a combination of near- and far-from resonant stroboSCAT probing conditions, one of which isolates heat alone, and calibration with steady state temperature-dependent reflectance contrast spectroscopy, we observe and quantitatively discern transient thermal and electronic contributions to the photogenerated dynamics. This new strategy enables isolation and characterization of the excitonic dynamics without concern for mistaking thermal contributions for electronic ones.
This capability with few-layer \ch{MoS2} complements the capabilities of microTRPL, which is largely restricted to monolayer TMDC measurements \cite{kulig_exciton_2018, zipfel_exciton_2020, goodman_substrate-dependent_2020} because additional layers lead to very low PLQY due to the emergence of an indirect band gap. Furthermore, we corroborate our results with a spatiotemporal model for heat and exciton populations. With the ability to detect temperature elevations as low as 100 mK, our study suggests that even a modest temperature elevation has a substantial effect on the optical response in few-layer \ch{MoS2}. More broadly, this work establishes a strategy for isolating electronic dynamics and transport in a wide range of conventional and emerging semiconductors that offers great potential for more incisively investigating thermal management and thermoelectric energy conversion.


\begin{figure}
    \centering
    \includegraphics[width=\linewidth]{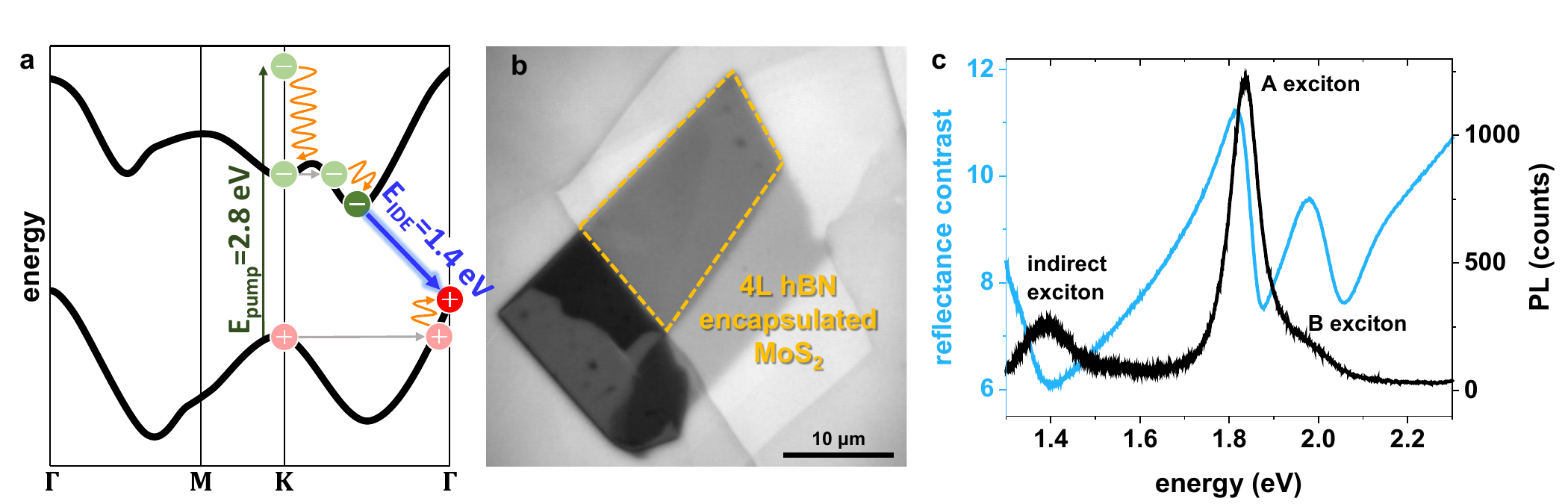}
    \caption{(a) Calculated single particle band structure for 4L \ch{MoS2} from Reference \citenum{splendiani_emerging_2010} showing the above-band gap pump excitation (green arrow) at the $K$-point of the Brillouin zone, \cite{mak_atomically_2010} followed by fast thermalization (orange) and intervalley scattering (gray) to the lowest energy indirect exciton (blue arrow). (b) Optical reflectance image of the entire \ch{MoS2} flake with the measured 4L hBN-encapsulated region outlined in yellow. (c) Steady state photoluminescence (PL) spectrum (black) showing strong emission from the direct A exciton near 1.77 eV with a blue shoulder corresponding to emission from the direct B exciton. Emission from the indirect exciton (1.4 eV) is comparatively weaker. The reflectance contrast spectrum (blue) exhibits dispersive resonance features near the PL peaks.}
    \label{fig:MoS2-fig1}
\end{figure}

To demonstrate the capability of identifying and simultaneously tracking electronic and thermal energy evolution, we focus on a hBN-encapsulated four-layer (4L) \ch{MoS2} flake fabricated using the hot pick-up technique and supported by a glass substrate. \cite{purdie_cleaning_2018} Before proceeding with spatiotemporally resolved transport measurements, we pre-characterize the sample. The relevant portion of the electronic structure is shown schematically in Figure \ref{fig:MoS2-fig1}a. \cite{splendiani_emerging_2010} Optical reflectance microscopy readily identifies the encapsulated 4L region outlined in yellow in Figure \ref{fig:MoS2-fig1}b. Additional sample thickness characterization is found in Figure S1. The steady state photoluminescence spectrum in Figure \ref{fig:MoS2-fig1}c shows characteristic strong emission from the spin-orbit split A ($\sim$1.8 eV) and B ($\sim$2 eV) direct excitons, with a strong B:A photoluminescence intensity and dispersive resonance amplitude ratio in the blue reflectance contrast spectrum indicative of good sample quality. \cite{mccreary_-_2018} Photoluminescence due to recombination of the lowest-energy indirect exciton (IDE) in the near-infrared ($\sim$1.4 eV), indicated by the blue arrow in Figure \ref{fig:MoS2-fig1}a, is comparatively weak. 

We use stroboSCAT to directly visualize heat and exciton transport in the 4L \ch{MoS2}. stroboSCAT is a differential pump-probe technique that leverages the sensitivity of optical scattering to image energy migration. This sensitivity arises from homodyne detecting even the high angle scattered field via interference with a reference, in this case specularly reflected probe light. The physical process is similar to transient reflectance spectroscopy, except for the important difference that it takes place in a microscope with a wide-field -- and often highly off-resonant --  probe pulse capable of amplifying the signal that scatters off of a more localized pump-induced excitation distribution. \cite{delor_imaging_2020} These particular experiments have nanometer spatial precision and $\sim$200 picosecond time resolution. Two pulsed diode laser sources are directed into the sample with a high numerical aperture microscope objective that also collects the signal and directs it to a CMOS camera (Methods). First, a near-diffraction-limited focused pump pulse of 2.8 eV photons (green arrow in Figure \ref{fig:MoS2-fig1}a) generates a Gaussian spatial distribution of photoexcitations in the sample which are then probed after a controllable time delay by a single-wavelength widefield laser pulse of either 2.4 or 1.8 eV. Specifically in 4L \ch{MoS2}, the pump pulse generates excitations at the $K$-point of the Brillouin zone. \cite{mak_atomically_2010} Efficient sub-ps intervalley scattering (horizontal gray lines in Figure \ref{fig:MoS2-fig1}a) and $\sim$fs phonon-assisted relaxation to the band edge (orange in Figure \ref{fig:MoS2-fig1}a) occur within the $\sim$200 ps experimental instrument response function (IRF), and we therefore expect the IDE to be the dominant electronic excitation on the time scale of our measurements. \cite{shi_exciton_2013} A differential signal image is generated through the difference of the image at a time delay after the pump and an image taken without a pump, normalized to the latter. The contrast in the image, $(R_\text{pump on} - R_\text{pump off})/R_\text{pump off}$, is commonly referred to as $\Delta R/R$. Despite the fact that the 4L \ch{MoS2} is not luminescent, the presence of photoexcitations modifies the material’s local dielectric function, generating transient contrast that evolves as a function of space and time according to the quantity and location of decaying and diffusing photoexcitations. Energy population dynamics are described by integrated population decays as a function of time. More importantly, transport is characterized via the mean squared expansion (MSE) of the population in space as a function of time, $\text{MSE} = \sigma^2(t) - \sigma^2(0) = 2Dt$ where $D$ is the diffusivity and $\sigma(t)$ is the width of the Gaussian population at time delay, $t$, assuming a cylindrically symmetric distribution. In principle, any photoexcitation, e.g., charge carriers, excitons, phonons, may be detected in this way since the measurement observable, elastically backscattered light, does not rely on the material absorbing or emitting light at a particular frequency. Furthermore, near an absorption resonance, the sign of the differential contrast can be tuned above (bright) or below (dark) the baseline (gray) background. For a dispersive optical resonance, the transient response leads to oppositely signed  $\Delta R/R$ on either side of the resonance. \cite{wolpert_transient_2012} Heat, which also modulates electronic resonances, may also modify $\Delta R/R$.

To isolate the effect of heating on the optical response, steady state reflectance contrast (RC) spectra \cite{li_measurement_2014} are measured at a range of temperatures from room temperature to 90$^{\circ}$C, as described below and in the Methods. RC spectra are obtained by measuring reflected spectra from the sample atop of the substrate ($R$) and separately under the bare glass substrate ($R_0$) and calculating $RC = (R-R_0)/R_0$. RC spectra over the measured temperature range are presented in Figure \ref{fig:MoS2-fig2}a, showing the characteristic A and B exciton resonance peaks redshifting and broadening with increasing temperature. From these spectra, we identify two spectral regimes, near- and far-from resonance, which tune the relative contribution to the stroboSCAT signal from heat and excitons, enabling distinction between the two, as described below. (Although features from the shifting electronic resonances of the A and B exciton dominate the spectra in the visible regime, the lowest-energy IDE carries most of the excited state population on our measurement time scales.) We select probe energies for spatiotemporal imaging from available discrete laser diode sources indicated by the red (near resonant, 700 nm or 1.77 eV) and green (far-from resonant, 515 nm or 2.41 eV) vertical lines in Figure \ref{fig:MoS2-fig2}a.

\section{Results and analysis}

\begin{figure}
    \centering
    \includegraphics[width=\linewidth]{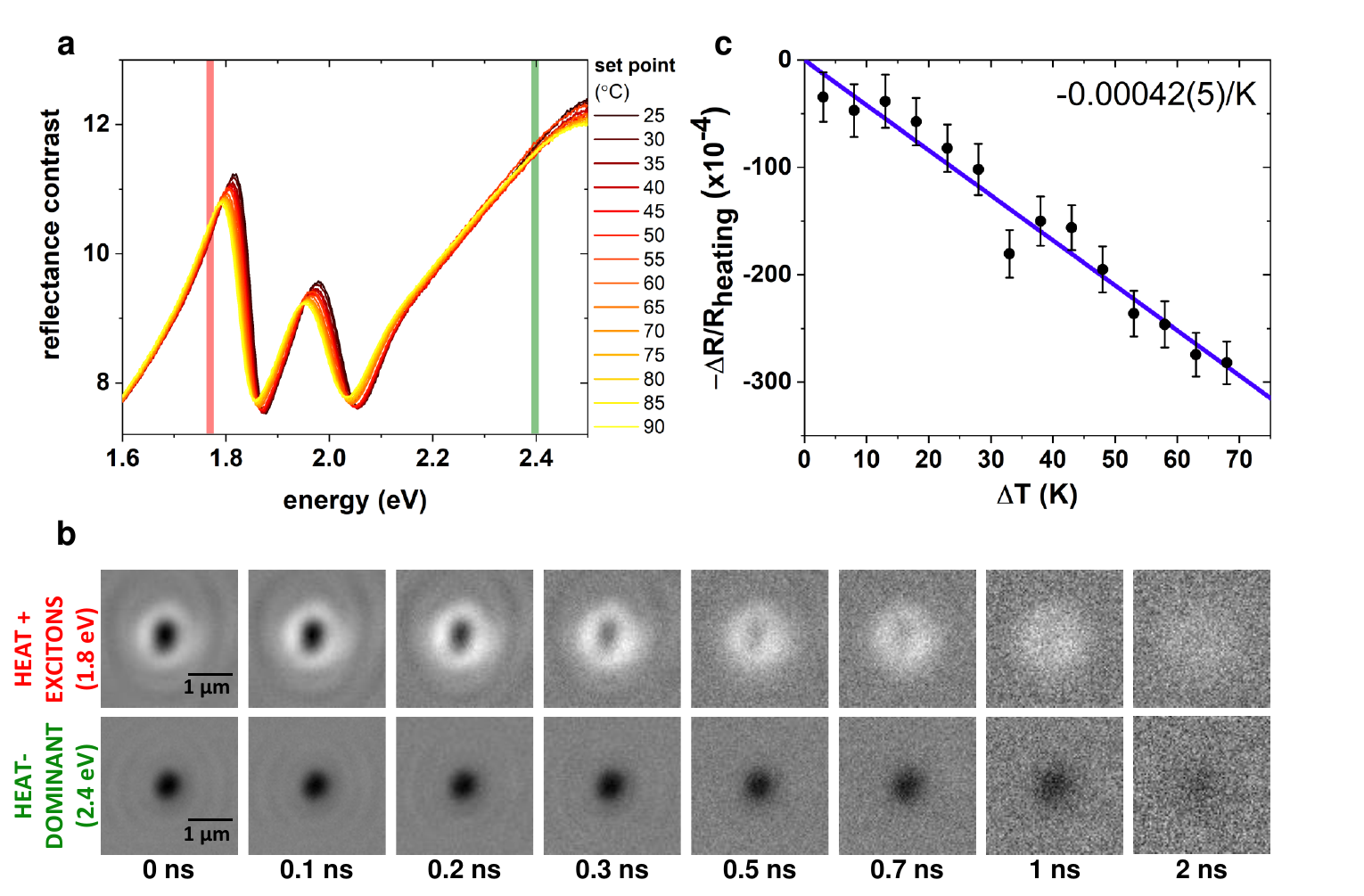}
    \caption{(a) Temperature-dependent reflectance contrast spectra over a range of temperature set points near the A and B exciton resonances. The near- and far-from-resonant probe energies are indicated with the red and green lines, respectively. (b) stroboSCAT time series captured with a near-resonant (top) and far-from-resonant (bottom) probe. The focused 2.8 eV pump generates a peak initial exciton density of $3.5 \times 10^{13}$ cm$^{-2}$. (c) Expected differential contrast due to heating, $\left( \frac{\Delta R}{R}\right)_\text{heating} = \frac{RC_\text{hot} - RC_\text{room temp}}{RC_\text{room temp}}$, at 1.77 eV. The vertical axis is multiplied by -1 for direct comparison to widefield stroboSCAT measurements. The linear fit (blue line) has a fixed intercept through the origin. Error bars include contributions from the propagated standard error of the mean from averaged spectra, calculated reflectance contrast over the 4 nm laser line, and fit error to the vertical intercept in Figure S7.}
    \label{fig:MoS2-fig2}
\end{figure}

In the same sample region, we collect two complementary stroboSCAT measurements over a 7 ns time window by probing at the near- and far-from resonant energies (Figure \ref{fig:MoS2-fig2}b). In both measurements, the same applied pump pulse fluence of 35 µJ/cm$^2$ generates an estimated peak exciton density of $3.5 \times 10^{13}$ cm$^{-2}$ (see Supporting Information), falling in an intermediate density regime where A-M interactions play a significant role but still being an order of magnitude below the Mott transition at which excitons would dissociate. \cite{chernikov_population_2015} Hot excitons thermalize and scatter to the indirect band edge within the IRF, transferring their excess energy to the lattice via efficient phonon emission. We therefore expect to observe both long-lived (several ns) IDEs and lattice heating simultaneously on our measurement time scales. In the far-from resonant stroboSCAT measurement, we observe negative (dark) contrast alone that decays and expands over several nanoseconds (Figure \ref{fig:MoS2-fig2}b, bottom), whereas in the near-resonant stroboSCAT measurement at the top of Figure \ref{fig:MoS2-fig2}b, we observe bright positive contrast beyond a dark, negative-contrast center that similarly decays over several nanoseconds, with positive contrast dominating after 1 ns. This contrast trend (negative contrast only probing at 2.4 eV, positive and negative contrast probing at 1.8 eV) and associated population dynamics and diffusivities persisted in multiple measured spots within the same sample and also in additional few-layer samples (Figure S3). We assume that each sign of contrast, positive or negative, in each measurement is generated by either heat or excitons, the two dominant forms of energy in the material following photoexcitation. Temperature-dependent RC spectra predict negative differential contrast due strictly to heating in the near-resonant measurement (Figure \ref{fig:MoS2-fig2}c and details below), therefore we deduce that positive contrast in the same measurement must be due to the presence of excitons. Using these assignments, we observe that excitons diffuse faster than heat, giving rise to positive amplitude extending beyond the heat-dominant negative contrast. This assignment is consistent with previous reports of exciton and heat diffusivity in \ch{MoS2}, in which exciton diffusivities are up to a few cm$^2$/s \cite{yu_giant_2020, uddin_neutral_2020} while reported heat diffusivities are slower at $\sim$0.2 cm$^2$/s. \cite{yan_thermal_2014} We note that these experiments are sensitive only to in-plane transport as the sample would need to be at least 30 nm, or $\sim$50 layers, thick in order to accumulate an interferometric phase flip that could indicate out-of-plane transport. \cite{delor_imaging_2020}

\begin{figure}
    \centering
    \includegraphics[width=\linewidth]{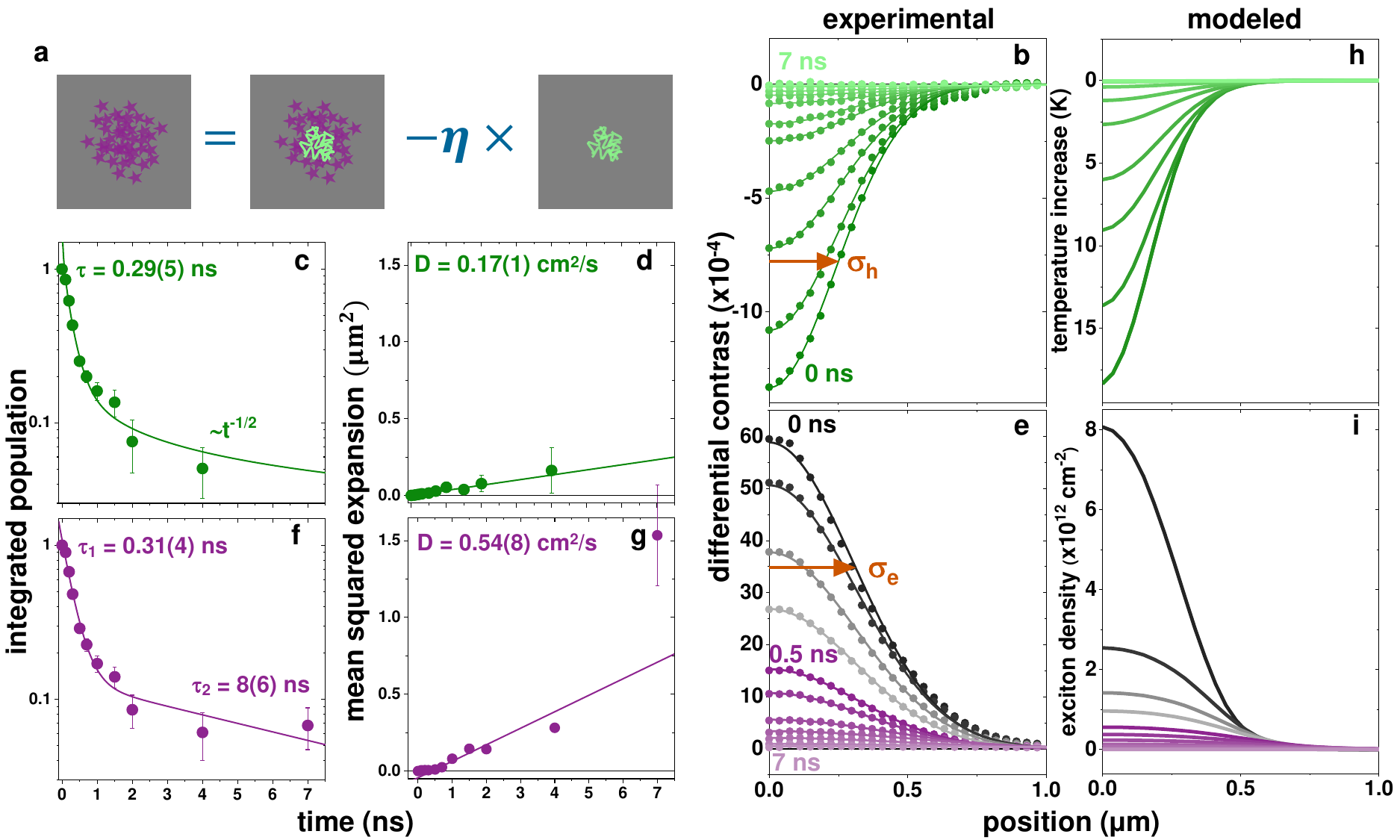}
    \caption{(a) Image arithmetic to isolate exciton distributions (purple) from thermal ones (green) when they coexist, with a scaling constant accounting for the wavelength-dependent sensitivity to optical perturbations depending on proximity to electronic resonances. (b) Azimuthal averages for the data shown in the bottom row of Figure \ref{fig:MoS2-fig2}b representing thermal distributions. (c,d) Integrated population decay fit to a single exponential plus power law (c) and mean squared expansion fit to a line (d) for the thermal distributions in (b). (e) Azimuthally averaged isolated exciton profiles after scaled image subtraction of the near- and far-from-resonant datasets in Figure \ref{fig:MoS2-fig2}b. (f,g) Integrated population decay fit to a biexponential (f) and mean squared expansion fit to a line starting at 0.5 ns (g) for the exciton distributions in (e). (h,i) Predicted thermal (h) and exciton (i) distributions from a spatiotemporal kinetic model best fit of the experimental data in (b,e). Orange arrows indicate the Gaussian fit widths to the time-zero population profiles for excitons and heat with $\sigma_\text{e}=0.3 $ µm and $\sigma_\text{h}=0.2 $ µm.}
    \label{fig:MoS2-fig3}
\end{figure}

To most readily compare the datasets at the two imaging wavelengths, we apply a pixel size correction to account for their different point spread functions (PSFs) (Figure S4). With this correction, the spatial extent of the positive signal measured with the near-resonant probe is demonstrably larger than the corrected negative signal measured with the far-from resonant probe (Figure S5). This observation suggests that unique dynamics give rise to the differing spatial extent of the positive and negative contrast signals respectively probed near and far from resonance, furthermore confirming that they represent distinct photoexcited species. As the positive contrast has already been assigned to an exciton population, we deduce that the negative contrast far-from resonance is dominated by heat. Based on the probe spectral proximity to electronic resonances in the system, we estimate that the excitonic contribution far-from resonance is suppressed by a factor of 25 relative to the near-resonant exciton contribution. Furthermore, a single Gaussian function fits all far-from resonant data well, therefore we deduce that the measurement is dominated by the optical response due to heating, and any potential contribution from excitons (positive or negative) is below our detection limit. It is therefore possible to isolate the positive exciton contribution to the near-resonant dataset despite its spatial overlap with the negative heat contribution by subtracting the far-from resonant heat-dominant measurement using ``image arithmetic'' (Figure \ref{fig:MoS2-fig3}a). This strategy requires quantitative knowledge of the difference in strength of the optical response due to heating at the two probe energies, the magnitude of which depends on the proximity to exciton resonances. We introduce a scaling factor, $\eta$, into this image subtraction to account for this difference: $\eta = \frac{\text{strength of optical response due to heat at 1.8 eV}}{\text{strength of optical response due to heat at 2.4 eV}}$. The far-from resonant stroboSCAT measurement characterizes the optical response due to heating at 2.4 eV. To quantitate the optical response due only to heating near resonance at 1.8 eV, we refer to the temperature-dependent reflectance contrast spectra in Figure \ref{fig:MoS2-fig2}a. To reframe the RC in units of stroboSCAT contrast, we calculate the difference in reflectance contrast at 1.8 eV found at each given elevated temperature and the value measured at room temperature and normalize it to the room-temperature response, $\Delta R/R_\text{heating}$ (Figures S4 and S5). The result is plotted in Figure \ref{fig:MoS2-fig2}c for each heater set point up to 90$^{\circ}$C ($\Delta T = 70$ K) with a linear fit, the slope of which, -0.00042(5)/K, quantifies the predicted differential stroboSCAT contrast associated with a given temperature increase. To estimate the temperature at which to compare to $\eta$, we use a spatiotemporal kinetic model, described in more detail below, to iteratively fit exciton and heat experimental profiles until the predicted maximum achieved temperature in the \ch{MoS2} matches the value of $\eta$ used in the fit. With this method we estimate that the sample reaches a maximum temperature of 18 K above room temperature after accounting for fast interfacial heat transfer to surrounding hBN, a reasonable estimate that is close to the predicted temperature increase from thermalization to the band edge (see Supporting Information). The scaling factor $\eta$ is therefore given by the ratio between the predicted maximum stroboSCAT contrast due to heat near resonance divided by the maximum stroboSCAT contrast at time zero in the far-from resonant measurement: $\eta = \frac{-0.00042(5)/\text{K} \times 18 \text{ K}}{-0.00124(2)} = 6.1(4)$.


With the interpretation and strategy developed above, we obtain the population dynamics and transport parameters for both heat and excitons in the encapsulated 4L \ch{MoS2}. We determine the thermal dynamics by fitting each azimuthally averaged spatial distribution of each time point in the far-detuned stroboSCAT measurement to a Gaussian function (Figure \ref{fig:MoS2-fig3}b). We find that the integrated temperature profile has an initial fast $\sim$300 ps decay due to interfacial transfer to the encapsulating hBN, and then heat transfers more slowly, limited by the rate of heat diffusion in the hBN (Figure \ref{fig:MoS2-fig3}c). Although hBN is a good thermal conductor, its capacity to sink heat generated in \ch{MoS2} is limited by the evolving temperature gradient between the two materials and the finite volume of hBN, and since the amplitude of the Gaussian temperature distribution in the hBN drops due to lateral heat diffusion as $t^{-1/2}$, this scaling determines the rate-limiting thermal transfer for the \ch{MoS2} in the $\sim$ns time frame (see Supporting Information). We obtain an in-plane heat diffusivity in the \ch{MoS2} of 0.17 cm$^2$/s, consistent with the reported lateral thermal conductivity for \ch{MoS2} (Figure \ref{fig:MoS2-fig3}d). \cite{yan_thermal_2014}

In order to quantitatively discern exciton dynamics from heat dynamics, we developed a careful strategy of frame-by-frame azimuthal profile subtraction of the near- and far-from resonant stroboSCAT datasets (Figure \ref{fig:MoS2-fig3}a). First, each Gaussian fit profile representing the heat population (temperature profile) in the far-detuned measurement is width- and amplitude-adjusted by a time-dependent PSF correction factor. This operation generates the \textit{shape} of the isolated heat distribution that would have been measured with the near-resonant probe. Next, we multiply the PSF-corrected heat profiles by the above-deduced scaling factor, $\eta$, to quantitatively represent the differential contrast profile due to heating near resonance (Figure S5). Finally, we azimuthally average the total near-resonant stroboSCAT signal and subtract the PSF-corrected and scaled thermal component obtained from the far-from resonant dataset. The isolated radial exciton profiles are near-Gaussian, as shown in Figure \ref{fig:MoS2-fig3}e. The spatially integrated exciton population dynamics fit best to a biexponential exciton decay, which we attribute to density-dependent A-M interactions that dominate at early time delays when exciton densities are higher ($\tau_1 = 310$ ps), followed by slow nonradiative recombination of the IDE over $\tau_2 \sim 8$ nanoseconds (Figure \ref{fig:MoS2-fig3}f). The exciton profiles at time delays earlier than 500 ps (grayscale) do not appear to expand, presumably due to the pump-induced profile and its early time changes being beneath the spatiotemporal resolution of these measurements. The extracted exciton diffusivity for $\geq$500 ps, once it is possible to see the profiles expanding, is 0.5 cm$^2$/s, in agreement with other measurements in few-layer TMDCs (Figure \ref{fig:MoS2-fig3}g). \cite{ganatra_few-layer_2014, liu_direct_2020} We note that excluding the final 7 ns data point from the linear fit does not change this result within the fitting error. Repeating this dynamical analysis over a range of values of $\eta$ from 1.4 to 7 enables an estimate of the uncertainties in the extracted diffusivities, based on uncertainty in the initial maximum sample temperature elevation that we can safely bound between 4 and 20 K.

We support these findings with a spatiotemporal kinetic model that describes the coupled dynamics of excitons and heat. Additional details are included in the Supporting Information. We developed a simple set of coupled equations to capture the expansion and decay of excitons whose energy is overwhelmingly (due to very low PLQY) converted to heat via either nonradiative hot carrier relaxation following optical excitation or A-M decay: \cite{kulig_exciton_2018, perea-causin_exciton_2019}

\begin{equation} \label{exc-evolution}
    \dot{N}(r,t) = D_\text{X}(r,t) \nabla^2 N(r,t) - \frac{1}{\tau_\text{X}}N(r,t) - R_\text{A-M}N^2(r,t) + G(r,t)
\end{equation} \label{heat-evolution}
\begin{equation}
    \dot{T}(r,t) = \alpha N^2(r,t) - \frac{1}{\tau_\text{T}}[T(r,t) - T_0] + D_\text{T} \nabla^2T(r,t) + \beta N(r,t) + \gamma G(r,t)
\end{equation}

\noindent Equation \ref{exc-evolution} describes the evolution of the exciton population, $N(r,t)$. The first term describes exciton diffusion due to the pump-induced exciton population gradient, with an exciton diffusivity of $D_\text{X}$. The second term describes single exciton recombination, where $\tau_\text{X}$ is the recombination lifetime. The third term describes biexciton recombination due to A-M interactions, where $R_\text{A-M}$ is the A-M coefficient. Finally, $G(r,t)$ describes the generation of excitons from the pump pulse, which has a temporal pulse width of 72 ps and a spatial width, $\sigma$, of 168 nm. Equation \ref{heat-evolution} describes the temperature or heat population, $T(r,t)$. The first term accounts for the temperature increase due to nonradiative relaxation of hot excitons that are created via A-M recombination, where $\alpha = \frac{R_\text{A-M}E_\text{G}}{c}$, $E_\text{G}$ is the indirect band gap energy, and $c$ is the specific heat. The second term describes the decay of the temperature profile back to its initial room temperature value, $T_0$, with a lifetime $\tau_\text{T}$, and the third term describes heat diffusion with diffusivity $D_\text{T}$. The fourth term describes the temperature increase due to nonradiative single-exciton recombination, where $\beta = \frac{(1-\text{PLQY}) E_\text{G}}{\tau_\text{X}c}$. The last term describes the heat generated when excitons relax to the band edge after pump excitation, where $\gamma = \frac{E_\text{P}-E_\text{G}}{c}$ and $E_\text{P}$ is the pump pulse energy. In addition, when propagating these equations in time, the heat decay phenomenologically transitions to a $t^{-1/2}$ scaling after 700 ps to represent thermal transfer to hBN discussed in the Supporting Information. To fit the model to the time series of exciton profiles extracted above, along with the corresponding heat profiles obtained far-from resonance, we allow $\tau_\text{X}$ and $D_\text{X}$ to vary only within the experimentally determined uncertainty and $R_\text{A-M}$ to vary within the literature estimates.\cite{uddin_universal_2021} We approximate the indirect band gap by the peak position of the IDE PL (1.4 eV). All other parameters are fixed by our experimentally obtained values or from literature values.\cite{volovik_enthalpy_1978, mak_atomically_2010}

Figure \ref{fig:MoS2-fig3}h,i and Figure S8 show excellent qualitative agreement between the modeled and experimentally obtained exciton and thermal profiles. The model predicts an initial maximum exciton density of $\sim 8 \times 10^{12}$ cm$^{-2}$, lower than our experimental estimate based on the pump fluence and sample absorbance. We expect this discrepancy arises because the model accounts for the finite pump duration and subsequent exciton decay occurring within the experimental IRF. (If we suppress A-M interactions, exciton decay, and exciton expansion in the model, we recover the initial condition for $N(0,0) = 3.5 \times 10^{13}$ cm$^{-2}$.) The predicted maximum temperature at experimental time-zero, (predominantly from exciton thermalization after above band gap excitation), is 18 K, which is the value that is consistent with $\eta$ of 6.1 (see Supporting Information). The best fit $R_\text{A-M}$ coefficient, a parameter we cannot constrain with our experiments due to the experimental IRF, is $2.6 \times 10^{-3}$ cm$^{-2}$, which is lower than for monolayer TMDCs, but still within the expected range for multilayer TMDCs. \cite{sun_observation_2014, zipfel_exciton_2020, uddin_universal_2021} Overall, the model supports our experimental finding that excitons diffuse slightly faster than heat, importantly enabling spatial differentiation between the two, which we described as instrumental in the differential contrast assignment.

\section{Discussion}

Having described our observations and analysis strategies and corroborated our results with simulations, we turn to a discussion of our findings and of what they reveal and implicate. Below, we first establish the consistency of our physical model with other transient microscopy results that indirectly reveal a role for heat to impact exciton populations in 2D TMDCs. Second, we explore stroboSCAT’s favorable temperature sensitivity, which, under the conditions employed in this work, is as good as $\sim$100 mK. Third, we explore the value in employing our spatio-spectro-temporal approach and discuss complementary, related strategies that could be used as a general toolkit to draw from, depending on the particular photophysical details of any given material, given the ubiquity of heat generation. Finally, we suggest strategies to treat increasingly complex combinations of energy carriers, and we point toward the utility of our approach in elucidating and leveraging new mechanisms of electron-phonon coupling in thermal management and thermoelectrics.

To test our model-related findings, we note that Perea-Causín et al. \cite{perea-causin_exciton_2019} also employed transient optical microscopy to study the interplay between heat and excitons in 2D TMDCs, albeit with different experimental parameters. When registering the exciton photoluminescence of monolayer \ch{WS2}, following ultrafast pump excitation leading to a far greater nonequilibrium-exciton effective temperature elevation, they observed remarkable ``halo-like'' spatial photoluminescence profiles, which they attributed to radially outward excitonic transport driven by strong temperature gradients. To relate our model to these results, we added the Seebeck term $\frac{\sigma S}{q}\Delta T$, as in Perea-Causín et al., \cite{perea-causin_exciton_2019} to Equation \ref{exc-evolution} to reflect that exciton diffusivity is driven not only by exciton density gradients but also by thermal ones. 
Here, $\sigma(r,t) = N(r,t)q\mu$ is the electric conductivity for exciton density $N$, elementary charge $q$, and exciton mobility, $\mu$. $S$ is the Seebeck coefficient, an intrinsic material parameter. By using the Perea-Causín parameters into the model, we indeed generate halo-shaped excitonic profiles (Figure S9). If, however, we retain the Seebeck term to attempt to fit our experimental results that have a less impulsive pump pulse and furthermore allow $\eta$ to vary, we find that an exceedingly large Seebeck coefficient ($S>10,000$ V/K) would be required to generate halo-like exciton profiles. For this reason, we rule out this Seebeck regime to model our observations in 4L \ch{MoS2}.
This finding is also consistent with results from Zipfel et al. \cite{zipfel_exciton_2020} who also do not observe exciton halos in encapsulated samples with $R_\text{A-M} \lesssim 10^{-3}$ cm$^{-2}$.

\begin{figure}
    \centering
    \includegraphics[width=\linewidth]{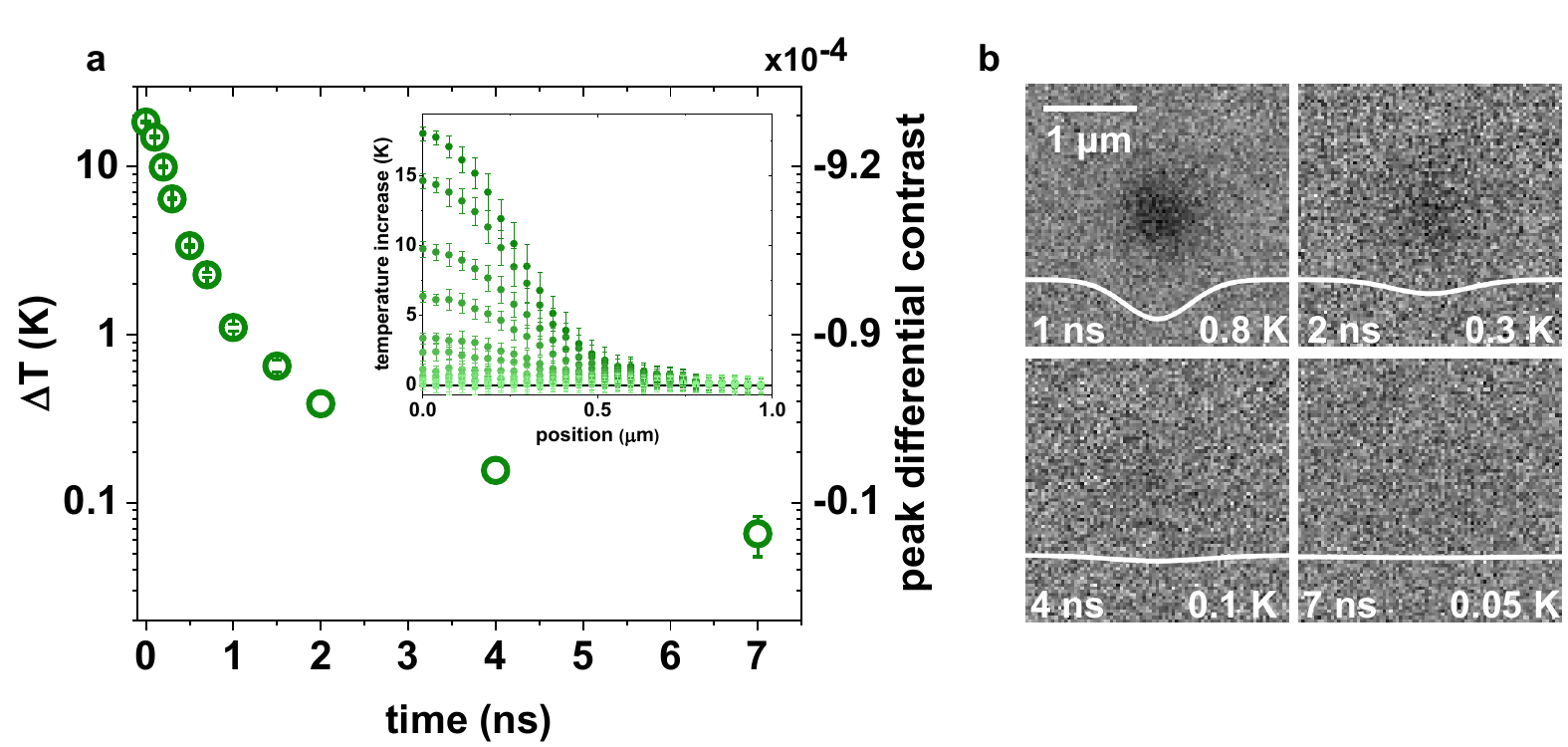}
    \caption{(a) Gaussian amplitude fit decay for the far-from resonant probe dataset. The vertical axis is calibrated using the maximum predicted sample temperature elevation from the spatiotemporal model (18 K) and the measured maximum differential contrast value. The inset shows the azimuthal averages in Figure \ref{fig:MoS2-fig3}b with a rescaled vertical temperature axis. Error bars represent the standard error of the mean of the azimuthal averaging. (b) Far-from resonant probe images labeled with the maximum temperature achieved in each frame. All images are shown on the same contrast scale. White traces are Gaussian fits to the raw azimuthally averaged data.}
    \label{fig:MoS2-fig4}
\end{figure}

We also wish to highlight a valuable byproduct of developing thermal sensitivity capabilities in a transient optical microscopy – namely that stroboSCAT can serve as a highly sensitive non-contact thermometry with excellent spatial resolution compared to infrared analogs. For example, based on the 18 K temperature elevation calculation in the 4L \ch{MoS2}, we reframe in Figure \ref{fig:MoS2-fig4} the relaxation of the temperature in Figure \ref{fig:MoS2-fig2}b and Figure \ref{fig:MoS2-fig3}f to establish the sensitivity of this thermometry. Figure \ref{fig:MoS2-fig4}a relabels the axis of Figure \ref{fig:MoS2-fig3}f with the time-dependent peak temperature at the center of the heat distribution. The radial profiles of corresponding time delays, also in Figure \ref{fig:MoS2-fig3}b, are shown as an inset, and some example $\Delta R/R$ images are included in Figure \ref{fig:MoS2-fig4}b. At the longer time delays, we thus establish the ability to resolve $\sim$0.1 K temperature elevation. Although it is difficult to directly compare with the metrics of more conventional thermal imaging, \cite{adhikari_photothermal_2020} this value seems on par with it and enjoys substantially higher spatial resolution. While the present data were collected by only averaging for 7 minutes per time point, we estimate, based on the 12-bit CMOS camera well depth and the shot noise limit, that our sensitivity should indeed be in the range of 100 mK. With increased averaging, microscope stabilization, and detector sensitivity, this limit could be pushed into the tens of mK regime, making stroboSCAT an exquisite thermometry approach with added high spatial resolution. Not only is this newfound spatially-resolved sensitivity to temperature very powerful for discerning heat and charge in photoexcited materials, but it should find great utility in thermal management characterization in the semiconductor device sector.

Regarding the toolkit that we have developed and its more general applicability, our spatio-spectro-temporal approach is able to characterize the complex, overlapping electronic and thermal system dynamics even though their contributions would be difficult to disentangle from spectroscopic data alone; discerning these dynamics in 4L \ch{MoS2} was possible despite an absence of strongly distinct time scales in the excitonic and thermal population dynamics. Another strategy to achieve the same result could be to use the temperature-dependence of the reflectance contrast to calibrate the response to heat at 2.4 eV, not only at 1.8 eV, where the proximity to a zero-crossing precluded this process in the current work. Careful selection of another far-from resonance probe wavelength could enable this alternative strategy for future work. As a general strategy, for each specific sample, one must carefully consider the best approach to isolate the electronic contribution to the photoexcited signal based on spectral information, as discussed here, and the extent to which the thermal and electronic diffusivities create a separation of time scales, which is minimal here but was, for example, sufficient to investigate silicon.\cite{delor_imaging_2020} Furthermore, there are generally multiple different processes by which heat is generated that can occur on different time scales relative to the generation and evolution of electronic excitation. Regardless, the impact of heat on transient measurements is substantial in many materials, not only for few-layer TMDCs. \cite{lukianova-hleb_influence_2009, smolin_distinguishing_2018, cooper_physical_2018} Any type of non-radiative process, from above gap excitation, to nonlinear processes like annihilation, including Auger-Meitner effects, to ‘standard’ non-radiative decay, will generate heat. By measuring the photoinduced response in a reflectance geometry, the thermal response is revealed most clearly due to its superior sensitivity to the real part of the dielectric function, relative to transmission-based (or photoluminescence) measurements. Again, while it has been painstakingly investigated in semiconductor transient spectroscopy, \cite{ruppert_role_2017, cooper_physical_2018} which is largely performed in reflectance geometry due to the opacity of semiconductors in the visible and near-infrared parts of the spectrum, the higher sensitivity and the addition of the spatial variable that stroboSCAT affords provide additional helpful constraints to discern heat and charge. The spatial coordinate is helpful not only because of access to instantaneous spatial distributions but also to the time rate of change in the spatiotemporal evolution, which yields transport parameters such as diffusivity. \cite{ginsberg_spatially_2020} We therefore anticipate this newfound ability to characterize the coexistence, transport, and interplay of heat and charge in materials to be highly general and to enable a more detailed and reliable mechanistic understanding of a material’s physical properties and phenomena far more broadly than in TMDCs alone.

Regarding strategies for the future, the possibility exists to interrogate materials with more than two types of energy carriers, for example, free carriers, trions and excitons, coexist together with one another and also with heat. 
In 4L \ch{MoS2}, we probed sufficiently far from a zero crossing in the transient reflectance spectrum in all cases to avoid sign-changes in the differential contrast associated with each type of energy. In materials with additional, distinct electronic species, the combination of measuring at additional pump energies and fluences to tune the relative densities of distinct photoinduced energy carriers could enable them to be distinguished. Furthermore, a continuously tunable probe source could be leveraged to increase the signal-to-noise ratio of photoexcitations that may not appreciably modify the local dielectric function. \cite{su_dark-exciton_2022, xu_ultrafast_2022} 

It should also be possible to measure both electronic and thermal energy transport in more heterogeneous material configurations as are found in various devices, especially when composed of organic semiconductors. Due to stroboSCAT's differential nature, structural irregularities are effectively normalized away, leaving behind a clear view of how transport is impacted by them. If there are localized and well-separated sub-diffraction defects or interfaces, their impacts on transport can be resolved, as we saw previously in Ref. \citenum{delor_imaging_2020}, generating steps in the mean squared expansion. This can be employed to reveal anomalous diffusion of heat\cite{utterback_nanoscale_2021} as well as charge. When defects or interfaces are more closely spaced, the observed transport properties will reflect the net effect of the constituent materials and interfaces. Nevertheless, transport measurements on individual material components and on more controlled interfaces may be used as control experiments to aid in explaining the aggregate heterostructure behavior. Furthermore, the impact of out-of-plane heterogeneities may be resolved by investigating phase changes in the stroboSCAT contrast or by considering the population dynamics observed in the presence of interfaces in addition to the lateral transport, similar to the way that we treat thermal transfer between \ch{MoS2} and hBN in this work or to the way that interfacial transfer is examined in Ref. \citenum{utterback_nanoscale_2021}. 

We also envision a range of additional utilities of stroboSCAT in elucidating mechanisms of electron-phonon coupling and deepening our understanding of intrinsic thermal–electronic energy conversion and transport. Spatiotemporally monitoring charged photoexcitations and phonons simultaneously opens new doors for discovering mechanisms of electron-phonon scattering. In particular, understanding nonradiative decay pathways facilitated by traps, interfaces, defects, A-M interactions and natural background doping will elucidate design principles for engineering higher PLQY materials and directed or enhanced diffusion lengths. \cite{lu_bandgap_2014, wang_ultrafast_2015, li_defect-mediated_2019, li_phonon-suppressed_2019} Characterizing the potential interplay between heating and electronic energy flow could inform thermal management strategies by revealing the dominant factors and mechanisms that tune electron-phonon coupling. Furthermore, while thermal management aims to mitigate the impact of unwanted and deleterious heat dissipation, for example in semiconductor electronics, stroboSCAT also has the capability to directly measure transport anisotropies and thermoelectric effects in which heat is harnessed to do useful electronic work. For example, directly measuring the intrinsic Seebeck coefficient, an important factor in the intrinsic figure of merit for thermoelectrics, across different device configurations or material thicknesses may address challenges in efficiently upcycling heat loss through conversion to electricity or otherwise controlling heat flow in operating devices that suffer from poor performance due to self heating. \cite{dresselhaus_low-dimensional_1999} In particular, the interferometric depth-dependent stroboSCAT contrast in sufficiently thick layered van der Waals materials might be able to distinguish between out-of-plane transport and more rapid in-plane transport facilitated by strong covalent bonding, a potentially important design parameter for thermoelectric devices.

\section{Conclusion}

In conclusion, we use a combination of optical scattering microscopy and temperature-dependent reflectance contrast spectroscopy to co-measure and discern the photoinduced dynamics of heat and excitons in 4L \ch{MoS2}. This capability is a generalizable consequence of stroboSCAT’s unique spatially-, spectrally-, and temporally-resolved contrast mechanism that is sensitive to any perturbation that modifies a material’s local dielectric function. The spatiotemporal energy maps play a key role in identifying overlapping energetic populations with distinct contributions to the differential contrast. Our results agree with previous characterizations of few-layer \ch{MoS2}, are robust to experimental uncertainties in the estimated sample temperature elevation, and demonstrate a temperature sensitivity as good as 100 mK, ushering in a new era for spatiotemporally-resolved optical microscopy to discern charge and heat and their potential interplay. \cite{varghese_pre-time-zero_2023} With quantitative energy-carrier-specific tracking down to few-ps time scales, provided structural heterogeneities are sufficiently spaced for the distinct transport properties they produce to be spatially resolved, directly characterizing and explaining the factors that give rise to the optoelectronic properties of a wide range of emerging semiconducting materials, including intrinsic thermoelectrics and low-dimensional or organic electronic devices, is now possible without having to rely on complex models or assumptions.

\section{Materials and Methods}

\subsection{\ch{MoS2} preparation and characterization}

We use the hot pick-up technique to fabricate hBN-encapsulated few-layer \ch{MoS2} heterostructures on coverglass. \cite{purdie_cleaning_2018} Briefly, hBN and \ch{MoS2} are exfoliated onto 285 nm of thermally oxidized \ch{SiO2} on Si. We make stamps consisting of PDMS covered with a thin film of the thermoplastic polymer polycarbonate (PC). Using these stamps, we first pick up the top hBN, then the desired \ch{MoS2} flake, and then the bottom hBN, all at 50$^{\circ}$C. We deposit the stack onto \#1.5 coverglass, which serves as the substrate for all stroboSCAT measurements, at 180$^{\circ}$C, and then dissolve the PC in chloroform.

We characterize the hBN-encapsulated \ch{MoS2} sample using optical microscopy, Raman spectroscopy, and atomic force microscopy (AFM). The 4L \ch{MoS2} flake measured in the main text is outlined in blue in Figure S1a,b. The AFM image in Figure S1b shows that our sample has large ($>7\times7$ µm), homogeneous, bubble-free areas, which are ideal for measuring in stroboSCAT. The separation between the two \ch{MoS2} Raman peaks demonstrates that these samples are 4 layers thick Figure S1c,d. \cite{lee_anomalous_2010}. The hBN layer between the \ch{MoS2} flake and glass substrate is thin enough ($\sim$5 nm) to allow easy optical access to the \ch{MoS2} layer within the 1.4 NA objective’s depth of field.

\subsection{stroboSCAT measurements}

stroboSCAT measurements were performed at room temperature using three Picoquant pulsed laser diode sources, one to excite the sample at 440 nm (LDH-D-C-440) and two probe sources at 515 nm (LDH-D-C-520) and 700 nm (LDH-D-C-705). The base laser repetition rate was set to 10 MHz with the pump modulated at 660 Hz. Pump-probe time delays were electronically controlled by the laser driver with $<$20 ps precision. The pump and probe beams were spatially filtered through 25 and 50 µm pinholes, respectively, before being combined with a dichroic mirror (DMLP505, Thorlabs) and directed into the objective of an inverted microscope stage with a 50/50 beamsplitter. The pump was focused onto the sample with this high numerical aperture oil-immersion objective (Leica HC PL APO 63x/1.40NA) to a FWHM of $\sim$300 nm. The probe was focused into the back focal plane of the same objective to enable widefield illumination in the sample plane. The backscattered light was collected through the same objective, filtered (ThorLabs FB520-10 or Chroma ET720/60m) to reject the pump light, and focused onto a CMOS detector (PixeLINK PL-D752) using a 500 mm focal length imaging lens, resulting in an overall image magnification of 157.5$\times$. The instrument response function was measured to be $\sim$240 ps, primarily limited by the pulse duration of the diode sources (60-110 ps). Extensive details of the stroboSCAT technique are described in previous work. \cite{delor_imaging_2020}

Differential $\Delta R/R$ stroboSCAT images at each probe time delay are constructed from the difference between a widefield image collected after a pump pulse excitation and a widefield image collected without pump pulse excitation, normalized to the latter: $\Delta R/R = (R_\text{pump on} - R_\text{pump off})/R_\text{pump off}$. Each of these sequentially-taken individual images is obtained within the 1.3 ms camera exposure. To construct the images in Figure \ref{fig:MoS2-fig2}c, a total of 3500 pairs of pump-on + pump-off image exposures at each time delay were collected in each of 10 scans over the full set of time delays. The results of each time delay from these 10 scans were averaged together for a total averaging time of 7 minutes per time-delayed differential image. Images are contrast-adjusted so that the grayscale baseline is in the center of the scale with the maximum signal magnitude defining both positive and negative scale bounds. The -5 ns time delayed image is subtracted from each subsequent time delayed image as a baseline correction. We performed similar measurements in the same sample $>$10 times at 5 different pump fluences (5-85 µJ/cm$^2$) yielding similar contrast and dynamical trends. Measurements in a separate 4L \ch{MoS2} sample also yielded similar results. In constructing these differential images with only 1.3 ms between each pump-on or pump-off image acquisition, many structural irregularities are effectively normalized away, leaving behind the well-defined edge features and homogeneous material regions imaged in this sample system.

\subsection{Temperature-dependent reflectance contrast spectroscopy}

Broadband emission from a stabilized tungsten-halogen lamp source (ThorLabs SLS201L) is spatially filtered through an optical fiber then focused into a 0.9 NA air objective in an Leica DMi8 inverted microscope, illuminating the sample within a $\sim$1 $\mu$m spot size. The reflected light output is fiber-coupled to a calibrated Princeton Instruments Spectrometer (HRS300) with $\sim$1.5 nm spectral resolution. Spectra were collected by averaging 500 frames (10 ms exposure per frame) together after heating and equilibrating the sample using a PID-controlled metal ceramic heater (ThorLabs HT19R).

\subsection{Spatiotemporal model}

A set of coupled equations representing exciton and heat dynamics and transport and their interchange was recast in natural units and solved using the \texttt{pdepe} function in MATLAB. Multistart optimization was performed with 500 starting points over a constrained 3-parameter space with least squares minimization to the measured heat and isolated exciton profiles. More details are provided in the Supporting Information.

\begin{acknowledgement}
We thank W. Tisdale and M. Delor for valuable discussions and input in the earlier stages of this work. stroboSCAT method development for this work has been supported by STROBE, A National Science Foundation Science \& Technology Center under Grant No. DMR 1548924. Sample preparation, stroboSCAT and all reflectance measurements, and all data analysis were supported by the “Photonics at Thermodynamic Limits” Energy Frontiers Research Center of the U.S. Department of Energy, Office of Basic Energy Sciences under award no. DE-SC0019140. Modelling was supported by the Center for Computational Study of Excited State Phenomena in Energy Materials (C2SEPEM) under the U.S. Department of Energy, Office of Science, Basic Energy Sciences, Materials Sciences and Engineering Division under contract no. DE-AC02-05CH11231, as part of the Computational Materials Sciences Program. H. L. Weaver acknowledges a National Science Foundation Graduate Research Fellowship (DGE 1106400).  D.J. acknowledges the support of the Computational Science Graduate Fellowship from the U.S. Department of Energy under Grant No. DE-SC0019323. N.S.G. acknowledges an Alfred P. Sloan Research Fellowship, a David and Lucile Packard Foundation Fellowship for Science and Engineering, and a Camille and Henry Dreyfus Teacher-Scholar Award.
\end{acknowledgement}

\begin{suppinfo}
Thickness characterization of encapsulated four-layer \ch{MoS2}; Additional stroboSCAT measurements in few-layer \ch{MoS2}; Transfer matrix calculations; Point spread function correction; Temperature-dependent reflectance spectroscopy and construction of differential contrast due solely to heating; Maximum sample temperature estimate from calorimetry; Long-time temperature decay scaling; Spatiotemporal model; Spatiotemporal kinetic model with Perea-Causín et al. experimental parameters
\end{suppinfo}

\bibliography{maintext_bib.bib}

\end{document}


\pagebreak
\section{1. Thickness characterization of encapsulated four-layer \ch{MoS2}}  
\begin{figure}
    \centering
    \includegraphics[width=0.8\textwidth]{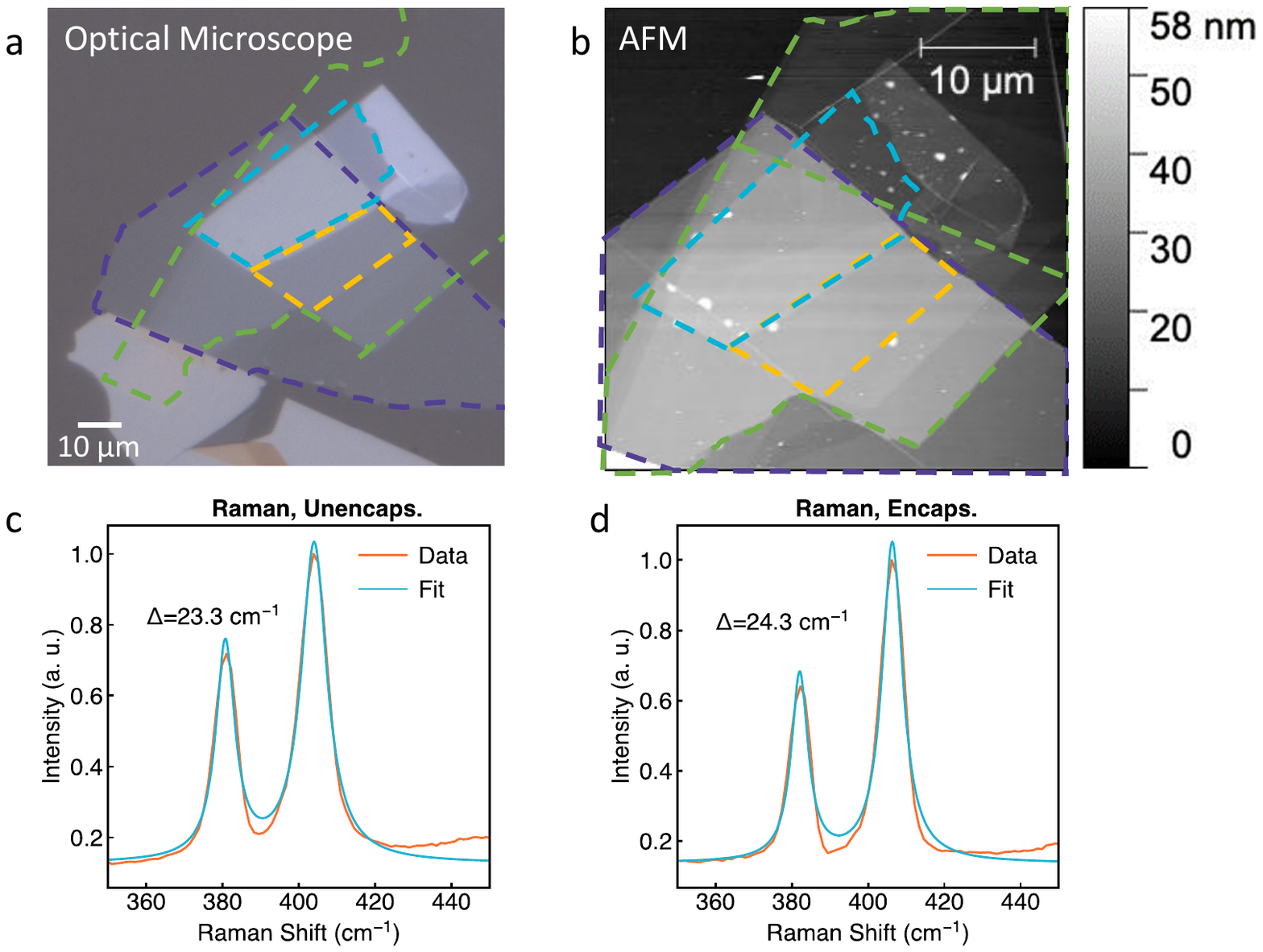}
    \caption{(a) Optical microscope image of hBN-encapsulated \ch{MoS2} where blue outline is 4L \ch{MoS2}, yellow outline is 1L \ch{MoS2}, green outline is substrate-adjacent hBN and purple outline is air-adjacent hBN. (b) Atomic force microscopy scan of sample used to identify hBN thicknesses.  (c, d) Raman spectroscopy of out-of-plane A$_\text{1g}$ mode in pre-encapsulated (c) and post-encapsulated (d) region, demonstrating a shift that corresponds to an \ch{MoS2} thickness of 4 layers.}
    \label{fig:sample-characterization}
\end{figure}

\section{2. Additional stroboSCAT measurements in few-layer \ch{MoS2}}  

\begin{figure}
    \centering
    \includegraphics[width=1\textwidth]{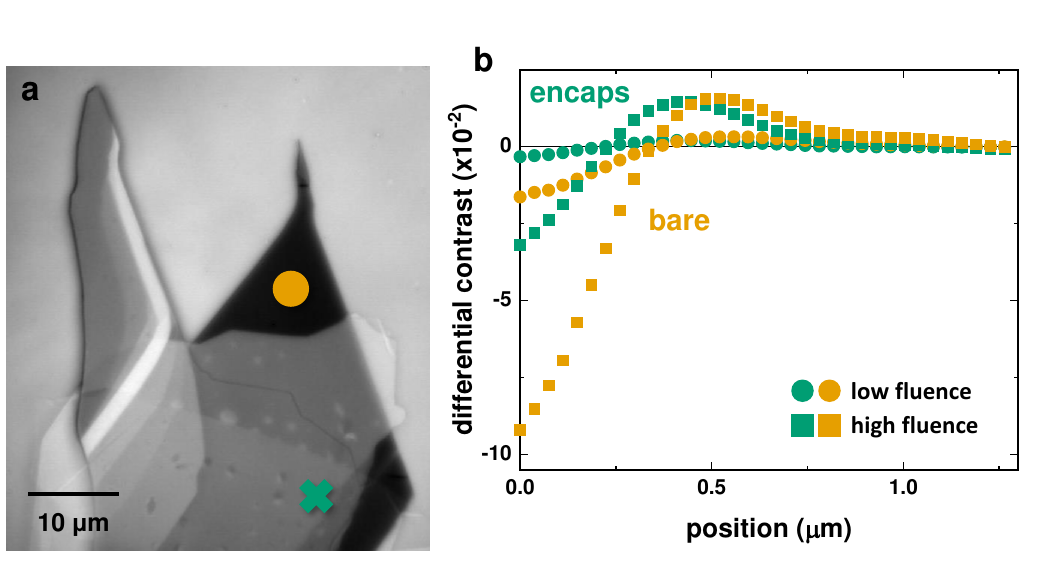}
    \caption{(a) Optical reflectance image of a separate few-layer \ch{MoS2} sample with an hBN-encapsulated region (near green ex) and unencapsulated region (near gold circle). (b) Time zero radial profiles measured in each region, with ``low fluence'' corresponding to an exciton density of $\sim 2 \times 10^{13}$ cm$^{-2}$ and ``high fluence'' corresponding to an exciton density of $\sim 10 \times 10^{13}$ cm$^{-2}$.}
    \label{fig:caps-vs-unencaps}
\end{figure}

We performed measurements in a separately prepared 4L \ch{MoS2} sample with and without hBN encapsulation (Figure S\ref{fig:caps-vs-unencaps}a). While the positive contrast excitonic contributions are similar in amplitude for the same excitation density, the negative contrast heat contribution is far more substantial in the unencapsulated \ch{MoS2}, owing to the high thermal conductivity (heat sinking) of the hBN. This data comparison allows us to change only the heat contribution. As a result, we observe that the net signal’s cross section in the figure changes from negative to positive values at a larger radius in the unencapsulated sample, as one would expect when adding a more substantial heat contribution to the same excitonic population (Figure S\ref{fig:caps-vs-unencaps}b).

\begin{figure}
    \centering
    \includegraphics[width=1\textwidth]{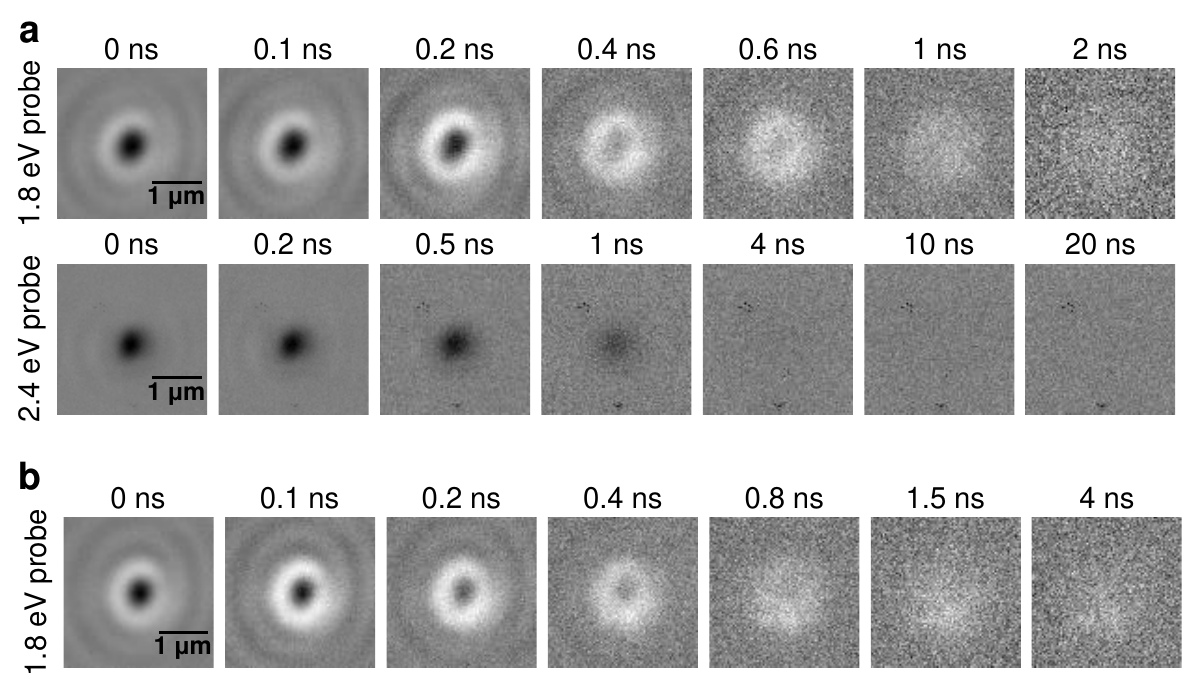}
    \caption{(a) Repeated stroboSCAT measurement at higher pump fluence (40 µJ/cm$^2$) in a different region of the sample measured in the main text. (b) stroboSCAT measurement at the encapsulated location marked in Figure S2a using the highest measured pump fluence (85 µJ/cm$^2$).}
    \label{fig:additional-measurements}
\end{figure}

We performed additional stroboSCAT measurements in 4L \ch{MoS2} at varying pump fluences from 5-85 µJ/cm$^2$ to confirm that the near- and far-from resonant contrast trends persisted (Figure S\ref{fig:additional-measurements}).

\section{3. Transfer matrix calculations}

We use transfer matrix calculations executed with the “tmm” Python software package \cite{byrnes_multilayer_2020, byrnes_tmm_nodate} to estimate the total absorbance over the four layers of  material in the sample, which is enhanced by multiple internal reflections and interference at the two hBN-\ch{MoS2} interfaces. We input complex refractive index values from the literature for 4L \ch{MoS2} \cite{song_layer-dependent_2019}  and a flat dispersion for hBN ($n_{\text{hBN}}=2.2$) with no absorbance in the visible ($k_{\text{hBN}}=0$) \cite{lee_refractive_2019}. The input layer thickness of \ch{MoS2} is 0.65 nm \cite{li_two-dimensional_2015}, while the hBN thicknesses are experimentally estimated with AFM (5 and 19 nm for the bottom and top layers, respectively). The calculation predicts that 6\% of incident photons are absorbed per layer, resulting in an overall 1/$e^2$ carrier density of 4.8$\times$10$^{12}$ cm$^{-2}$, or a peak carrier density of 3.5$\times$10$^{13}$ cm$^{-2}$. By contrast, a simple absorbance calculation using $\alpha(2.8 \text{ eV}) = 7.2 \times 10^5 \text{ cm}^{-1}$ yields an estimated 1/$e^2$ carrier density of 1.5$\times$10$^{12}$ cm$^{-2}$, or a peak carrier density of 9.3$\times$10$^{12}$ cm$^{-2}$, an underestimate by a factor of 3. The spatiotemporal model we describe in the main text and below estimates a maximum exciton density at time zero of $\sim$8$\times$10$^{12}$ cm$^{-2}$, lower than that predicted by the transfer matrix calculation because it takes into account a finite pump pulse duration (72 ps) over which some exciton-exciton annihilation occurs. Furthermore, a binding energy of $\sim 4 k_{\rm B} T$ (at room temperature) suggests that dissociated free carriers make up $< 2\%$ of the total photoexcited electronic population.

\section{4. Point spread function correction}

A diffraction-limited point spread function (PSF) is well-approximated by a normalized Gaussian function:

\begin{equation}
    \text{PSF}_\lambda (r) = \frac{1}{2 \pi \sigma^2_\lambda} \: \mathrm{e}^{-r^2/2 \sigma_\lambda}
\end{equation}

\noindent where $r^2=x^2+y^2$ and $\sigma_\lambda$ is given by the Abbe diffraction limit:

\begin{equation}
\label{eq:abbe_diff_lim}
    \sigma_\lambda = \frac{\lambda}{2 \text{ NA } 2\sqrt{2\ln{2}}}.
\end{equation}

\noindent In these experiments, excitons are generated by a nearly diffraction-limited 440 nm pump pulse and then efficiently convert to lattice heating, resulting in an initial Gaussian heat distribution:

\begin{equation}
\label{eq:pop_515}
    G_{\text{pop}}(r) = N_{\text{pop}} \: \mathrm{e}^{-r^2/2 \sigma_{\text{pop}}^2}
\end{equation}

\noindent where $\sigma_{\text{pop}}$ is the width of the actual heat population (temperature profile). This distribution is imaged in one set of experiments by a widefield 515 nm probe with a PSF given by

\begin{equation}
    \text{PSF}_{515}(r) = \frac{1}{2 \pi \sigma_{515}^2} \: \mathrm{e}^{-r^2/2 \sigma_{515}^2}
\end{equation}

\noindent where $\sigma_{515}$ is defined by Equation \ref{eq:abbe_diff_lim}. This imaging operation yields a measured Gaussian distribution, $G_{\text{meas,515}}(r)$, and may be represented by a 2D convolution:

\begin{equation}
    G_{\text{meas,515}}(r) = G_{\text{pop}}(r) \otimes \text{PSF}_{515}(r) = \frac{N_{\text{pop}} \sigma_{\text{pop}}^2 }{\sigma_{\text{pop}}^2 + \sigma_{515}^2} \: \mathrm{e}^{-r^2/2 (\sigma_{\text{pop}}^2 + \sigma_{515}^2)} = \frac{N_{\text{pop}} \sigma_{\text{pop}}^2 }{\sigma_{\text{meas,515}}^2} \: \mathrm{e}^{-r^2/2 \sigma_{\text{meas,515}}^2}
\end{equation}

\noindent where 

\begin{equation}
\label{eq:conv_thm_515}
    \sigma_{\text{meas,515}}^2=\sigma_{\text{pop}}^2 + \sigma_{515}^2.
\end{equation} 

From experimentally measured data at time-zero, $\sigma_{\text{meas,515}}=180$ nm, so for a diffraction-limited microscope, the initial heat distribution, from Equations \ref{eq:abbe_diff_lim} and \ref{eq:conv_thm_515}, has a width of $\sigma_{\text{pop}} = \sqrt{\sigma_{\text{meas,515}}^2 - \sigma_{515}^2} \approx 160$ nm, which is slightly larger than the pump width of $\sigma_{\text{pump}}\approx 130$ nm, to be expected since the heat distribution may expand slightly during the temporal overlap of the $\sim$100 ps pump and probe pulses.

If it were possible to measure only heat with a 700 nm probe, the measured Gaussian distribution, $G_{\text{meas,700}}(r)$, would be represented by a different convolution: 

\begin{equation} 
\label{eq:pop_700}
    G_{\text{meas,700}}(r) = G_{\text{pop}}(r) \otimes \text{PSF}_{700}(r) = \frac{N_{\text{pop}} \sigma_{\text{pop}}^2 }{\sigma_{\text{meas,700}}^2} \: \mathrm{e}^{-r^2/2 \sigma_{\text{meas,700}}^2}
\end{equation}

\noindent where 

\begin{equation}
\label{eq:conv_thm_700}
    \sigma_{\text{meas,700}}^2=\sigma_{\text{pop}}^2 + \sigma_{700}^2,
\end{equation}

\noindent and again, $\sigma_{\text{pop}}$ is the width of the actual generated heat population which, under identical excitation conditions, does not change with probe wavelength.

The goal is to transform the measured 515 nm heat distribution data to what would have been measured with a 700 nm probe in order to perform an accurate subtraction of the heat distribution from the total measured differential signal measured with 700 nm, which includes heat and excitonic contributions. The transformation of the measured 515 nm data for this purpose requires suppressing the amplitude and stretching the width of $G_{\text{meas,515}}(r)$ to represent the additional ``smoothing'' that would occur with a broader imaging PSF. To make this transformation more concrete, we define a correction factor, $\kappa$:

\begin{equation}
\label{eq:kappa}
    \kappa \equiv \frac{\sqrt{\sigma_{\text{pop}}^2 + \sigma_{700}^2}}{
    \sqrt{\sigma_{\text{pop}}^2 + \sigma_{515}^2}} = \frac{\sigma_{\text{meas,700}}}{\sigma_{\text{meas,515}}}
\end{equation}

\noindent where $\kappa >1$. We express Equation \ref{eq:pop_700} in terms of $\kappa$ and known variables. The amplitude of $G_{\text{meas,700}}(r)$ may be expressed as:

\begin{equation}
    \frac{N_{\text{pop}} \sigma_{\text{pop}}^2 }{\sigma_{\text{meas,700}}^2} = \frac{1}{\kappa^2} \frac{N_{\text{pop}} \sigma_{\text{pop}}^2 }{ \sigma_{\text{meas,515}}^2},
\end{equation}

\noindent and the Gaussian exponent of $G_{\text{meas,700}}(r)$ may be expressed as:

\begin{equation}
    -\frac{r^2}{2 \sigma_{\text{meas,700}}^2} = -\frac{1}{\kappa^2} \frac{r^2}{2 \sigma_{\text{meas,515}}^2} = -\frac{(r/\kappa)^2}{2 \sigma_{\text{meas,515}}^2} = -\frac{r'^2}{2 \sigma_{\text{meas,515}}^2}
\end{equation}

\noindent where $r = \kappa r'$ represents the ``contracted'' $r$-axis when imaging with 515 nm light, which must be multiplied by $\kappa$ in order to reproduce the effectively ``stretched'' $r$-dimension when imaging with 700 nm light. Therefore, to transform the measured 515 nm heat distribution to what would have been measured with 700 nm light requires dividing the measured 515 nm data everywhere by $\kappa^2$ and stretching the radial position axis by a factor of $\kappa$.

\begin{figure}
    \centering
    \includegraphics[width=1\textwidth]{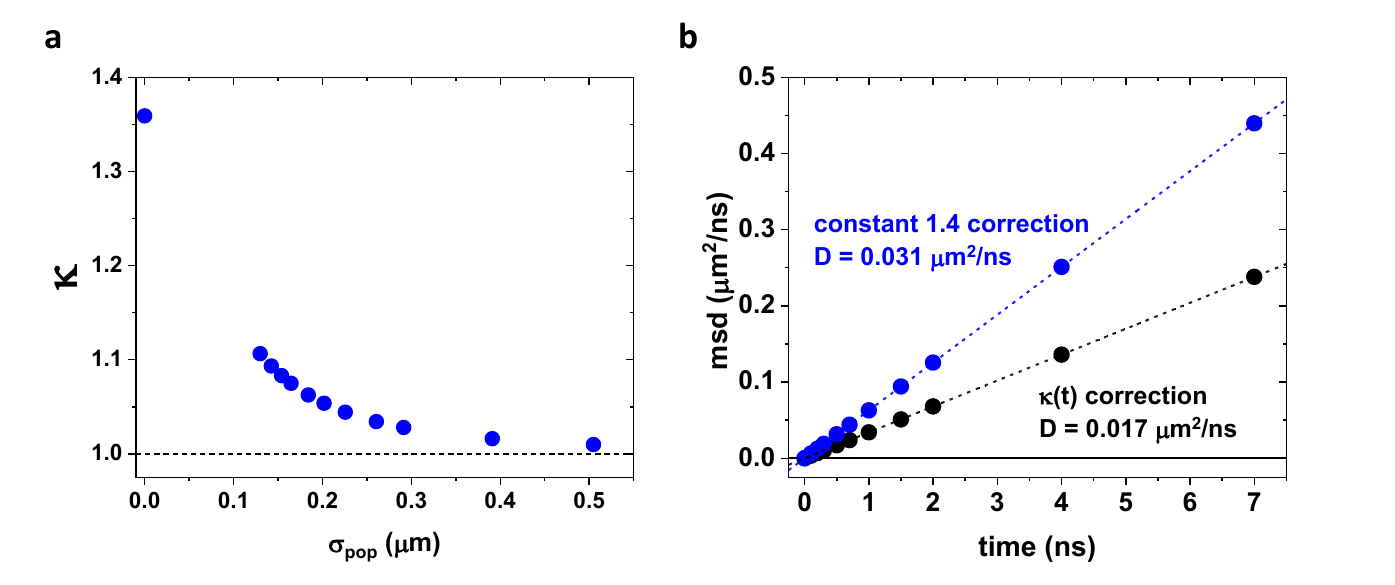}
    \caption{PSF correction. (a) Correction factor, $\kappa$, as a function of the actual population width, $\sigma_{\text{pop}}$. The maximum correction factor occurs for a delta function population and asymptotes to 1 (no correction) for increasingly broader populations, as expected. (b) Data transformation with a time-dependent correction factor $\kappa(t)$ (black) preserves the dynamics of the raw 515 nm data whereas a constant correction (blue) factor inflates the extracted diffusivity.}
    \label{fig:kappa}
\end{figure}

Because the heat distribution expands over time, the correction factor $\kappa$ is time-dependent:

\begin{equation}
    \kappa(t) = \frac{\sigma_{\text{meas,700}}(t)}{\sigma_{\text{meas,515}}(t)} = \frac{\sqrt{\sigma_{\text{pop}}^2(t) + \sigma_{700}^2}}{
    \sqrt{\sigma_{\text{pop}}^2(t) + \sigma_{515}^2}} 
\end{equation}

\noindent so that every time point must be amplitude- and width-corrected by a different (diminishing) factor (Figure \ref{fig:kappa}a). $\kappa(t)$ is calculated by: (1) extracting $\sigma_{\text{meas,515}}(t)$ from the Gaussian fit to the raw data at each time delay, (2) calculating $\sigma_{\text{pop}}(t)$ using (1) and Equation \ref{eq:conv_thm_515}, and (3) calculating  $\sigma_{\text{meas,700}}(t)$ using the result of (2) in Equation \ref{eq:conv_thm_700}. The time-dependent correction factors applied to the raw azimuthally averaged 515 nm data before subtracting from the raw azimuthally averaged 700 nm data are listed in Table \ref{tab:kappa}.

\begin{table}
\centering
\begin{tabular}{ c c }

 time (ns) & $\kappa$  \\ 
 0.0 & 1.078  \\  
 0.1 & 1.071 \\
 0.2 & 1.065 \\
 0.3 & 1.060 \\
 0.5 & 1.052 \\
 0.7 & 1.045 \\
 1.0 & 1.038 \\
 1.5 & 1.031 \\
 2.0 & 1.026 \\
 4.0 & 1.015 \\
 7.0 & 1.010 
\end{tabular}
\caption{Time-dependent PSF correction values \label{tab:kappa}}
\end{table}

Without accounting for the finite size of the imaged excited population, one may be tempted to use the ratio of the resolution limits at the two imaging wavelengths ($700/515 \approx 1.4$) as a global correction factor. However, the more complete picture described here demonstrates the importance of including the generated population's finite width and its time evolution in our analysis (Figure \ref{fig:kappa}a). We confirm that the time-dependent correction factor preserves the original measured dynamical information, including the mean squared expansion of the heat distribution, which should be invariant to imaging wavelength under identical excitation conditions (Figure \ref{fig:kappa}b).

\begin{figure}
    \centering
    \includegraphics[width=0.6\textwidth]{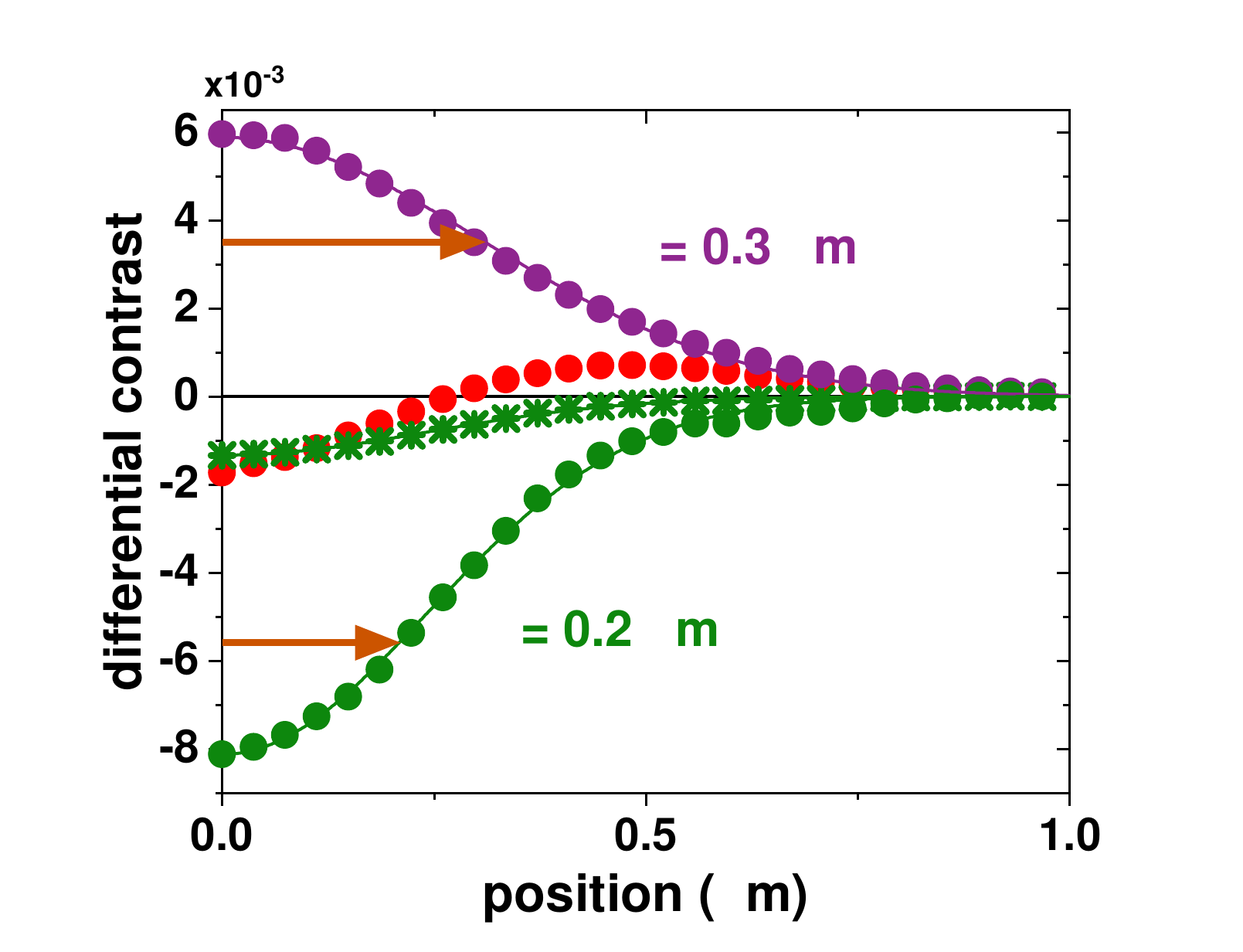}
    \caption{Time zero radial profiles of the 2.4 eV probe heat-dominated signal (green stars) following PSF correction, 1.8 eV probe raw signal (red), exciton profile (purple), and 1.8 eV heat profile (green circles).}
    \label{fig:prof_comp}
\end{figure}

The PSF-corrected radial profile for the 515 nm (2.4 eV, green circles) signal at time zero is shown in Figure \ref{fig:prof_comp} for comparison to the raw 700 nm (1.8 eV, red) signal and extracted exciton profiles (purple) generated from scaled subtraction of the PSF-corrected and $\eta$-scaled 515 nm data from the 700 nm data (green stars representing heat at 1.8 eV). The width of the corrected 515 nm signal is still much narrower than the full extent of the 700 nm signal, indicating that different imaging PSFs at the two probe wavelengths are not sufficient to explain the discrepancy. Instead, the population giving rise to the 515 nm signal diffuses more slowly during the instrument response time and must therefore be a different energy population than what gives rise to the faster-diffusing bright signal at 700 nm.

This experimental observation is corroborated by a calculation that compares the relative strength of the excitonic contribution to the optical response based on the relative proximity of the probe energy to its nearest electronic resonance. The dispersive lineshapes in the transient response follow a $1/(\omega - \omega_0)$ scaling detuned from resonant frequency $\omega_0$. The nearest electronic resonance to the 1.77 eV probe is at 1.79 eV (A exciton resonance) whereas the nearest electronic resonance to the 2.41 eV probe is at 2.07 eV (B exciton resonance). Furthermore, we estimate from the steady state photoluminescence spectrum in Figure 1c that the oscillator strength of the B exciton resonance is 10$\times$ smaller than that of the A exciton resonance. Therefore, the relative excitonic contribution to the far-from resonant measurement is suppressed by $\frac{1/(2.41-1.79)+0.1/(2.41-2.07)}{1/(1.79-1.77)+0.1/(2.07-1.77)} \approx 0.038$, or a factor of 25. We deduce that this negligible excitonic contribution is overwhelmed by the heat-induced signal probed at 2.41 eV. 

\section{5. Temperature-dependent reflectance spectroscopy and construction of differential contrast due solely to heating}

\begin{figure}
    \centering
    \includegraphics[width=1\textwidth]{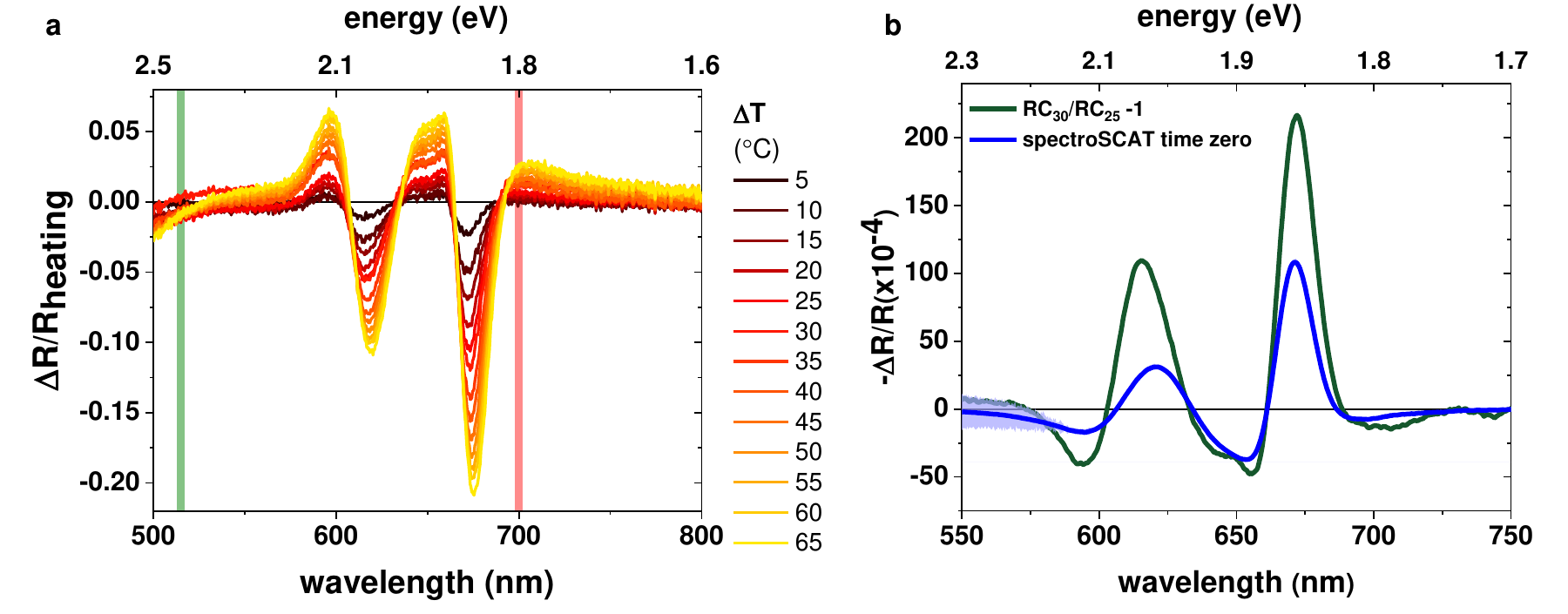}
    \caption{Reflectance spectroscopy as a function of temperature and compared to pump-induced differential reflectance (spectroSCAT). (a) Constructed differential reflectance spectra due to steady state heating normalized to the lowest controlled temperature set point spectrum (25$^{\circ}$C). A zero-crossing near 2.4 eV (green line) further complicates reliable quantification of the optical response due to heating at this energy while the trend near 1.8 eV (red line) is clear and monotonic. (b) Expected differential contrast due to steady state heating by $\Delta T = 5$ K (green) and time-zero photoexcited differential reflectance spectra (blue) averaged over a $\sim$500 nm line cut through the central excitation. Measurements are taken on the same 4L \ch{MoS2} sample in separate instruments in the Atwater and Ginsberg groups, respectively. Shaded error bars on the spectroSCAT curve are estimated from the fluctuations of the white light probe pulse near the second harmonic.}
    \label{fig:spectroscat}
\end{figure}

To identify the electronic resonances in our sample, we implement reflectance contrast spectroscopy. The same hBN-encapsulated 4L \ch{MoS2} sample that was measured with stroboSCAT was also measured in a separate commercial inverted microscope adapted for reflectance contrast spectroscopy. The use of an air objective in the reflectance contrast microscope adds an additional index-mismatched interface that enhances the overall reflection compared to what would be measured with an oil immersion objective. We use transfer matrix simulations, described in the above section, to estimate a correction factor to convert between the air and oil immersion objective cases: $R_{\text{oil}} \backsimeq R_{\text{air}} - \frac{R_{0,\text{air}}}{2} $ where $R$ is the reflectance under the sample and substrate and $R_0$ is the reflectance under the substrate only. The reflectance contrast, $RC = R/R_0$, may then be converted as $RC_{\text{oil}} = 2RC_{\text{air}} - 1$. 

The lowest controlled heater set point (25$^{\circ}$C) was used as a proxy for room temperature to normalize higher temperature set points in the construction of a differential reflectance spectrum due to sample heating:

\begin{equation} \label{eq:heating}
    \Delta R/R_{\text{heat}} = \frac{RC_{\text{hot}} - RC_{25 ^{\circ} \text{C}}}{RC_{25 ^{\circ} \text{C}}},
\end{equation}

\noindent yielding a qualitative representation of the optical material response due solely to heating (Figure \ref{fig:spectroscat}a). Poor lamp spectral power below 550 nm adds significantly to the error near the zero crossing and therefore precludes quantitative analysis of the expected $\Delta R/R $ contrast due to heating in this spectral regime. 

We note that the shape and location of resonance features in these differential heating spectra are strikingly similar to transient differential reflectance spectra measured after photoexcitation, suggesting that even modest sample heating of a few Kelvin dominates the transient response (Figure \ref{fig:spectroscat}b). We collect transient reflectance spectra by focusing a white light probe onto the sample in the stroboSCAT microscope and dispersing the reflected light through a home-built prism spectrometer. The white light is generated by focusing the fundamental output (1030 nm, 200 kHz) of a Light Conversion PHAROS ultrafast regeneratively amplified laser system into a 3 mm sapphire crystal. The excitation source is the same diode laser that was used for experiments in the main text. An external delay generator, triggered with the pulse output of the ultrafast laser, controls the electronic delays between pump and probe. White light fluctuations near the second harmonic of the fundamental add significant noise below $\sim$575 nm. 

\begin{figure}
    \centering
    \includegraphics[width=0.7\textwidth]{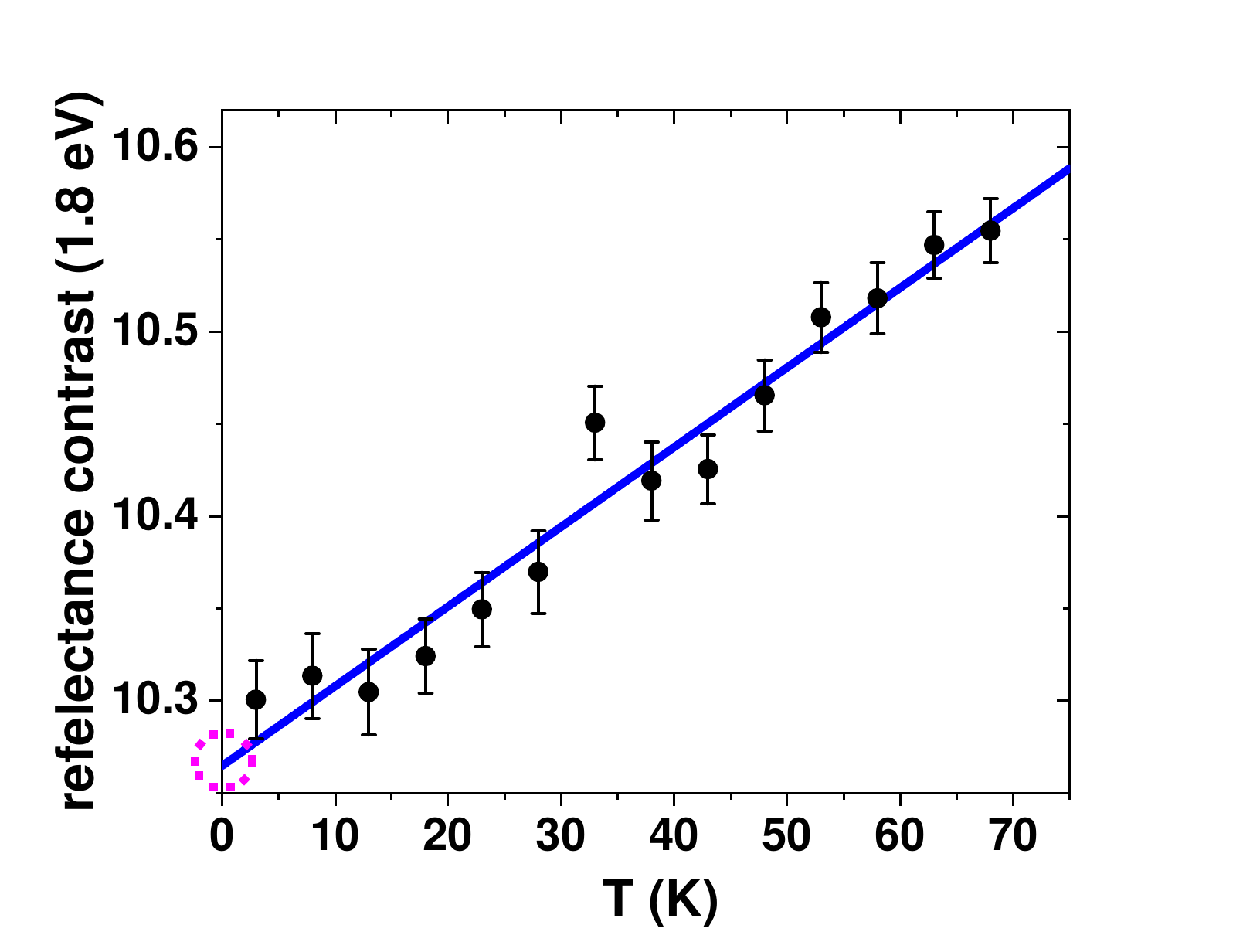}
    \caption{Reflectance contrast at 1.8 eV as a function of sample temperature elevation. The trend is extrapolated to room temperature ($\Delta T = 0$, pink dashed circle) using a linear fit (blue line). Error bars indicate the standard error of the mean propagated from averaging spectra and binning over the 4 nm laser line.}
    \label{fig:rc1.8}
\end{figure}

To estimate the temperature-dependent contribution to stroboSCAT contrast at the near-resonant 1.8 eV probe (Figure 2c in the main text), we first extrapolate the temperature-dependent reflectance contrast, binned over the 4 nm laser line, to room temperature (22$^{\circ}$C in the temperature-controlled stroboSCAT laser lab) and use this value to normalize the expected differential contrast due to heating as in Equation \ref{eq:heating} (Figure \ref{fig:rc1.8}). Finally, for direct comparison to stroboSCAT contrast, the result is multiplied by -1 to account for a relative difference of $\pi$ in the Gouy focusing phase of the focused and widefield probes used in the reflectance contrast and stroboSCAT measurements, respectively.

\section{6. Maximum sample temperature estimate from calorimetry}

We perform a simple calorimetry calculation to estimate the maximum temperature in the sample if all absorbed photons were converted to heat. Using a peak carrier density in the center of the excitation spot of $N_{\text{peak}} = 3.5 \times 10^{13}$ cm$^{-2}$, pump energy of $E_{\text{pump}} = 2.8$ eV and a heat capacity \cite{volovik_enthalpy_1978} of $c = 15.22$ cal/(mol$\cdot$K) $= 3.27 \times 10^{12}$ eV$\cdot$cm$^{-2}$K$^{-2}$, the maximum sample temperature increase is given by $\Delta T_{\text{max}} = N_{\text{peak}} E_{\text{pump}} c = 30$ K. Roughly half of this heat is due to thermalization to the indirect bandgap: $\Delta T_{\text{therm}} = N_{\text{peak}} (E_{\text{pump}} - 1.4 \text{ eV})c = 15$ K. 

\section{7. Long-time temperature decay scaling}

The thermal conductivity of hBN is 15$\times$ greater than in \ch{MoS2}, therefore we assume the heat transfer to hBN to be unidirectional and irreversible. Initially, rapid interfacial transfer from \ch{MoS2} to hBN occurs (few hundred ps), but once the temperature gradient has lessened, driving additional interfacial transfer is dependent on the time scale over which heat spreads throughout the hBN. If the very center ($r=0$) of the original exciton profile (and resulting) temperature profile in \ch{MoS2} is the most critical, we ask: How does the temperature of the hBN at $r=0$ scale as a function of time? To answer this question, we assume that (1) initially the temperature profile in hBN mimics that of the \ch{MoS2}, i.e., is Gaussian, and that (2) the temperature profile in hBN evolves according to the heat equation, i.e., standard diffusion with a mean squared expansion governed by $\sigma^2(t) - \sigma^2(0) \sim D_{\text{hBN}}t$. The amplitude of a two-dimensional Gaussian profile that expands due to diffusion is given by

\begin{equation}
    g(r=0) = \frac{1}{\sigma \sqrt{2 \pi}} \propto t^{-1/2}.
\end{equation}

Therefore, once initial heat transfer is limited by the rate at which heat diffuses in the hBN, the rate at which the temperature lessens in the hBN will also determine the rate of interfacial transfer and scale as $t^{-1/2}$. We could not fit the heat profiles without the introduction of this $t^{-1/2}$ decay.

We note that this sensitivity to the kinetic scaling of heat transfer tracks the full thermal population decay, enabling stroboSCAT to measure the efficiency of interfacial energy transfer between different contact materials, e.g., between the light-absorbing layer and an intervening layer on the glass substrate. \cite{utterback_nanoscale_2021}

\section{8. Spatiotemporal model}

Equations 1 and 2 in the main text are  recast in natural units and expressed in matrix form for the \texttt{pdepe} function in MATLAB:

\begin{multline}
    \frac{\partial}{\partial t} 
    \begin{bmatrix}
    u_1 \\
    u_2
    \end{bmatrix}
    = \frac{1}{r} \frac{\partial}{\partial r} \left( r 
    \begin{bmatrix}
    \mathrm{A} u_2 \partial_r u_1 + \mathrm{B} u_1 \partial_r u_2 \\
    \partial_r u_2
    \end{bmatrix}
    \right)  \\ + 
    \begin{bmatrix}
    - \frac{\tau_T}{\tau_X} u_1 - \tau_T R_\text{A-M} N_0 u_1^2 + \tau_T g - (\mathrm{A} + \mathrm{B}) \partial_r u_1 \partial_r u_2 \\
    \alpha N_0 \mathrm{C} u_1^2 - (u_2 - 1) + \beta \mathrm{C} u_1 + \gamma \mathrm{C} g
    \end{bmatrix}
\end{multline}

\noindent where

\begin{align}
  t' \equiv t/\tau_T \\
  r' \equiv r/\sqrt{D_T \tau_T} \\
  u_1(r',t') \equiv N(r,t)/N_0 \\
  u_2(r',t') \equiv T(r,t)/T_0 \\
  g(r',t') \equiv G(r,t)/N_0
\end{align}

\noindent and we define

\begin{align}
  \mathrm{A} \equiv \frac{\mu k_B T_0}{q D_T} \\
  \mathrm{B} \equiv \frac{q \mu s T_0}{D_T} \\
  \mathrm{C} \equiv \frac{N_0 \tau_T}{T_0}
\end{align}

The initial conditions are set to be $N(r,0) = 0$ and $T(r,0) = 300$ K. The boundary conditions are set so that the exciton and temperature fluxes go to zero. 

To model the pump pulse, which has a finite duration, we use a generating function, $G(r,t)$, which is a product of two Gaussian functions. The first is a Gaussian in space with a standard deviation of $\sigma_r$. At the center, the peak exciton density is $N_0$. The second is a Gaussian in time with a standard deviation of $\sigma_t$ and normalized to 1. Therefore,

\begin{equation}
    G(r,t) = N_0 \exp{\left[ \frac{-r^2}{2 \sigma_r^2} \right]} \, \frac{1}{\sqrt{2 \pi \sigma_t^2}} \, \exp{\left[ \frac{-(t-t_0)^2}{\sigma_t^2} \right]}
\end{equation}

\noindent or in natural units

\begin{equation}
    g(r',t') = \frac{1}{\tau_T \sqrt{2 \pi \sigma_t^2}} \, \exp{\left[ \frac{-r^2}{2 \sigma_r^2} \right]} \, \exp{\left[ \frac{-(t-t_0)^2}{\sigma_t^2} \right]}
\end{equation}

The exciton mobility and lifetime are allowed to vary over the range $15 < \mu < 22$ cm$^2$/V$\cdot$s and $1 < \tau_X < 23$ ns, constrained by the experimentally extracted diffusivity and long decay time constants across exciton profiles extracted from $1.4 < \eta < 7$. The A-M coefficient, $R_\text{A-M}$, is constrained from estimated literature values to $5 \times 10^{-5} < R_\text{A-M} < 5 \times 10^{-2}$ cm$^2$/s. The heat diffusivity and lifetime are fixed by the experimentally measured values. We estimate that, typical for few-layer \ch{MoS2} where light emission from the indirect band gap must be phonon-assisted, the photoluminescence quantum yield (PLQY) is $\sim$1\%, although we found that the peak time-zero temperature is insensitive to the value of the PLQY used. 

We used this spatiotemporal model to self-consistently determine $\eta$,

\begin{equation} \label{eq:eta}
    \eta = \frac{-0.00042(5)/\text{K} \times \Delta T_\text{max} }{-0.00124(2)}
\end{equation}

\noindent using the following process. We extracted exciton profiles for a given value of $\eta$, then ran the spatiotemporal model optimization on the experimental exciton and heat data. If the $\eta$ value predicted from the best fit maximum time-zero temperature, $\Delta T_\text{max,fit}$, did not agree with the value of $\eta$ used to generate the exciton profiles for the fit, then we extracted a new set of exciton profiles using the $\eta$ value predicted by $\Delta T_\text{max,fit}$ in Equation \ref{eq:eta} and performed the optimization again until the $\eta$ value predicted by $\Delta T_\text{max,fit}$ in Equation \ref{eq:eta} and the $\eta$ value used for the exciton profiles in the optimization agreed. 

\begin{figure}[H]
    \centering
    \includegraphics[width=1\textwidth]{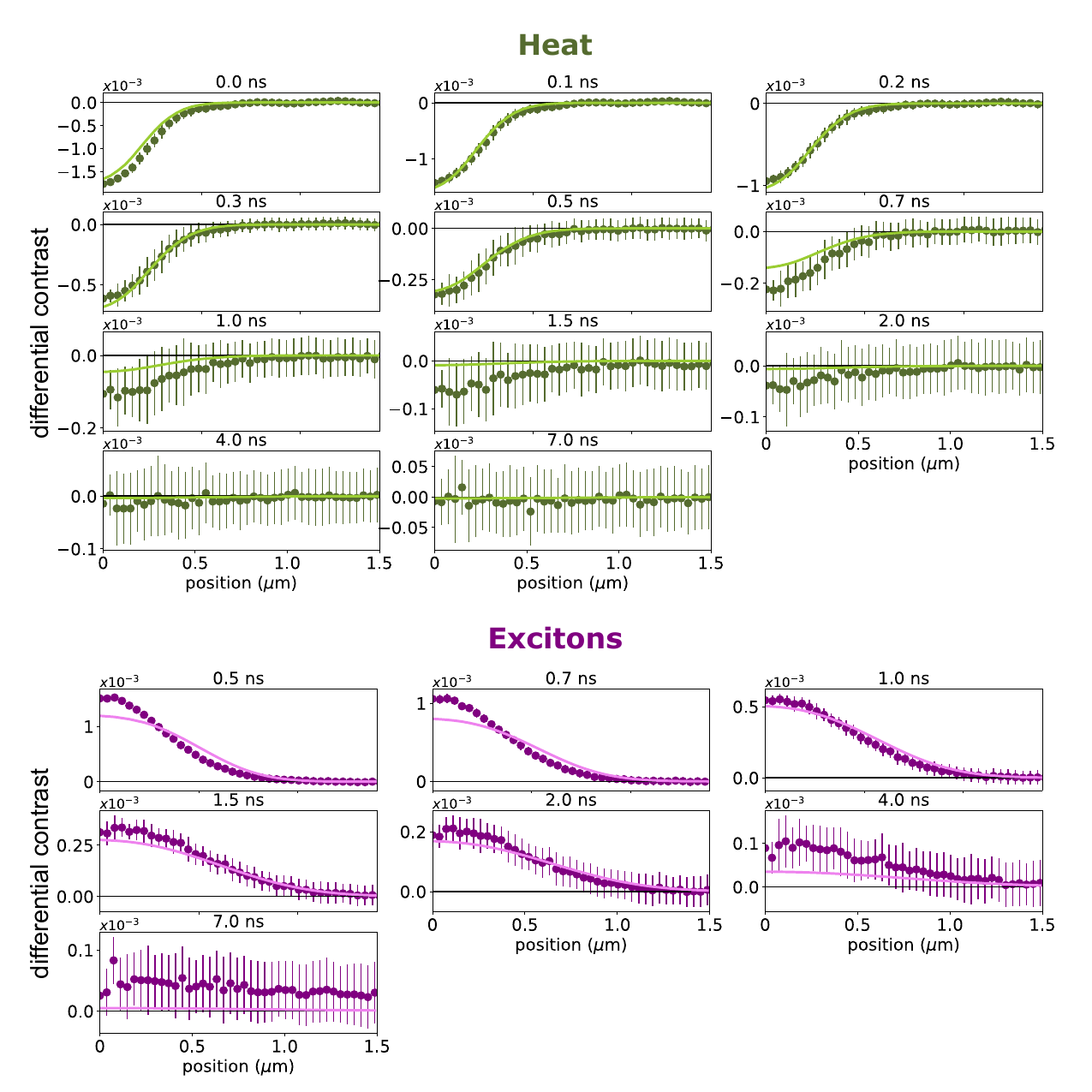}
    \caption{(top) Experimental far-from resonant heat profiles (dark green) with best fit from spatiotemporal model (light green). (bottom) Isolated experimental exciton profiles (dark purple) with best fit from spatiotemporal model (light purple).}
    \label{fig:model_on_exp}
\end{figure}

\section{9. Spatiotemporal kinetic model with Perea-Causín et al. experimental parameters}

Below are all model parameter values employed in two separate simulations. The right column summarizes the values either constrained or obtained via (experimentally bounded) fitting to the experimental stroboSCAT data. The middle column lists the corresponding parameters taken from Perea Causín et al.\cite{perea-causin_exciton_2019} to reflect that the model generates ``halo'' profiles consistent with their experimental observations using microTRPL.  

\begin{table}[H]
\centering
\begin{tabular}{ l c c }

   &  Perea-Causín et al.\cite{perea-causin_exciton_2019} & This work  \\ 
 $\tau_\text{X}$ [ns] & $0.7$ & $4.1$  \\  
 $D_\text{X}$ [cm$^2$/s] & $0.3$ & $0.57$ \\
 $R_\text{A-M}$ [cm$^2$/s] & $0.5$ & $0.00026$ \\
 $\sigma_\text{pump}$ [nm] & $174$ & $168$ \\
 $t_\text{pump}$ [ps] & $0.1$ & $72$ \\
 $E_\text{pump}$ [eV] & $2.43$ & $2.82$ \\
 $E_\text{BG}$ [eV] & $2.05$ & $1.38$ \\
 $N_\text{peak}$ [cm$^{-2}$] & $7 \times 10^{12}$ & $3.5 \times 10^{13}$ \\
 $c$ [J/g$\cdot$K] & $0.3$ & $0.4$ \\
 $\tau_\text{T}$ [ns] & $0.4$ & $0.3$ \\
 $D_\text{T}$ [cm$^2$/s] & $0.05$ & $0.2$ \\
 PLQY [\%] & $1$ & $1$  \\
$S$ [µV/K] & $1,000$ & $0$ 
 \end{tabular}
\caption{Spatiotemporal model parameters \label{tab:params}}
\end{table}

\begin{figure}
    \centering
    \includegraphics[width=0.75\textwidth]{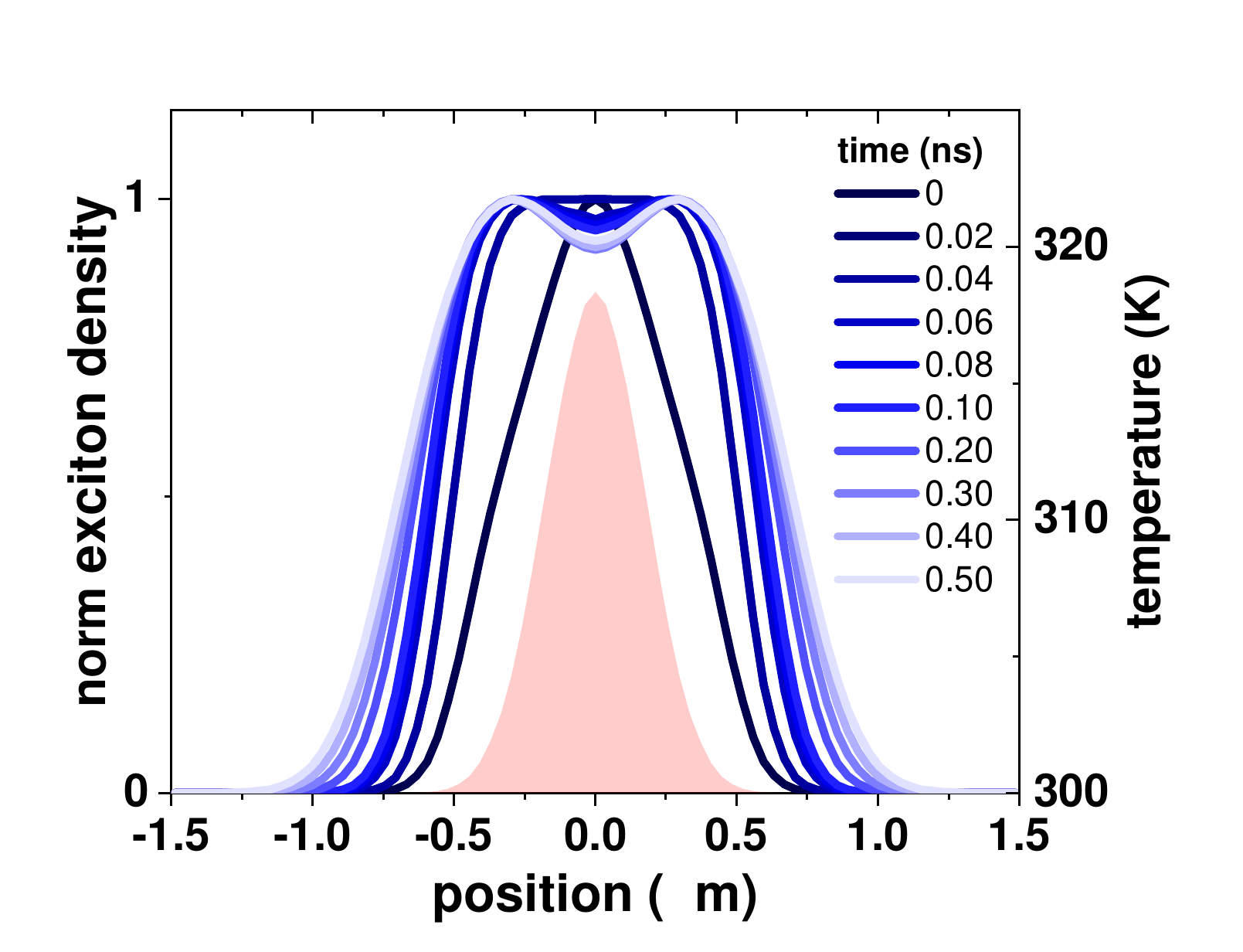}
    \caption{Normalized exciton profiles (blue) and time-zero temperature gradient (pink shaded curve) predicted with a spatiotemporal kinetic model that includes a Seebeck driving term in the exciton evolution equation over the same time delays measured in Figure 4 of Perea-Causín et al.\cite{perea-causin_exciton_2019}.}
    \label{fig:chernikov}
\end{figure}

\pagebreak
\bibliography{maintext_bib.bib}